\documentclass[onecolumn,usenatbib]{mn2e}

\usepackage{graphicx}
\usepackage{natbib}
\begin{document}

\title{Mountains on Neutron Stars: Accreted vs. Non-Accreted crusts} 

\author[B. Haskell et al]
{B. Haskell$^1$, D.I. Jones$^1$, N. Andersson$^1$  \\
$^1$ School of Mathematics, University of Southampton, 
Southampton, SO17 1BJ, United Kingdom}

\maketitle

\begin{abstract}
The aim of this paper is to compare the two cases of an isolated neutron star, with a non-accreted crust, and that of an accreting neutron star, with an accreted crust, and try to estimate which one of the two would make a better source of gravitational waves. In order to do this we must evaluate the maximum ``mountain'' that the crust can sustain in these two cases. We first do this using the formalism of \citet{ush} and find that the maximum quadrupole is very similar in the two cases, with the non-accreted crust sustaining a slightly larger mountain. We then develop a perturbation formalism for the problem, that allows us to drop the Cowling approximation and have more control over the boundaries. The use of this formalism confirms that there is not much difference between the two cases, but leads to results approximately one order of magnitude larger than those we obtain with the formalism of \citet{ush}.
\end{abstract}

\maketitle

\section{Introduction}

The possibility that the crust of a rotating neutron star may sustain a ``mountain'', thus creating a source of gravitational waves,  has recently sparked much interest. This is mainly due to the suggestion by \citet{bil}, echoing earlier ideas by \citet{wag} and \citet{pap}, that the gravitational wave torque due to such a mechanism may be dictating the spin equilibrium period for the LMXBs.
A recent estimate of how large a quadrupole one may expect was given by \citet{ush}, using the perturbation formalism of \citet{mac}. From their work it appears that the quadrupole may be big enough to balance the accretion torque, if the breaking strain of the crust (a very poorly constrained parameter) is taken to be near the maximum of the possible values.
Even though other gravitational wave mechanisms have been suggested for these systems (toroidal magnetic fields were suggested by \citet{cutler} and r-modes by \citet{rmode,rmode2}), and even though one can build models of the interaction between the accretion disc and the magnetic field that give the correct distribution of spin frequencies, as in \cite{us}, the matter is still of great interest.

There are in fact many questions that can still be asked, the main one being if there is a difference between the ``mountain'' that one can build on the non-accreted crust of a rotating neutron star and the one that can be built on the accreted crust of a neutron star in a binary system (such as the LMXBs). We shall see that these two different scenarios can lead to significantly different compositions for the crust.
The case of isolated neutron stars is particularly interesting as current gravitational wave detectors are beginning to put astrophysically  relevant limits on the ellipticities of known pulsars (\cite{ligo}).
In this work we will discuss this issue, and present a new scheme for calculating the maximum quadrupole which allows us to have better control of the boundary conditions at the base of the crust than in the scheme of \citet{ush}.
Another important point is that we do not adopt the Cowling approximation, which, as we will show, can lead to significant differences in the results.
We intend to present a formalism in terms of perturbations of a spherically symmetric crust, without making the Cowling approximation  that was made in \citet{ush}.

We will consider two cases for the crust core interface: in one case we will solve the hydrostatic equilibrium equations in the crust and in the core, and impose continuity of the tractions at the junction; in the other case we consider the core to remain unperturbed, and thus spherical. This is not physical but, as we shall see, allows us to estimate the errors associated with neglecting the perturbations of the core. The main point, however, is that if we leave the core unperturbed, this allows us to solve the TOV equations to obtain it's structure; we can thus obtain an estimate of how important relativistic effects may be. Such effects are relevant because, even though the Newtonian approximation is excellent at the low densities of the crust, the structure of the core can be highly relativistic and have an impact on global parameters such as the mass and radius of the star. 
We shall first develop the formalism for an $n=1$ polytrope and then apply it to the case of a realistic equation of state, for an accreted and a non-accreted crust.

\section{Maximum Mountain and Constant Strain in the Crust} 

In this section we shall briefly review the method of \citet{ush}, who were the first to perform a detailed calculation to estimate the maximum size of a ``mountain'' that a neutron star crust could sustain. As it will be necessary to refer to it, let us recapitulate briefly their work.

Let us assume that the crust responds elastically to pressure and density perturbations. We can thus treat it as a solid with a shear modulus $\mu(r)$ and write the stress tensor as:
\begin{equation}
\tau_{ab}=-pg_{ab}+t_{ab}
\end{equation}
where $g_{ab}$ is simply the flat 3-metric and $t_{ab}$ is the shear stress tensor of the crust, which vanishes in the fluid interior. We will consider the equilibrium shape of the star to be spherical, thus effectively representing the deformation of a non-rotating star, and treat $t_{ab}$ as a first order quantity, so that we have:
\begin{equation}
\delta\tau_{ab}=-\delta{p}g_{ab}+t_{ab}
\end{equation}
The equations we need to solve for equilibrium are thus:
\begin{equation}
\nabla^a\delta\tau_{ab}=\delta\rho g(r) r^a+\rho\nabla^a\delta\Phi
\label{elas}\end{equation}
where now $\rho$ indicates the background density and $g(r)=\nabla\Phi=GM/r^2$.
If we expand the perturbation in tensor spherical harmonics, as in \citet{ush}, we have:
\begin{eqnarray}
t_{ab}=t_{rr}Y_{lm}(r_ar_b-\frac{1}{2}e_{ab})+t_{r\perp}(r)f_{ab}+t_{\Lambda}(r)(\Lambda_{ab}+\frac{1}{2}Y_{lm}e_{ab})
\label{spando}
\end{eqnarray}
where
\begin{eqnarray}
\beta&=& \sqrt{l(l+1)}\\
e_{ab}&=&g_{ab}-r_ar_b\\
f_{ab}&=&\beta^{-1}r(r_a\nabla_bY_{lm}+r_b\nabla_aY_{lm})\\
\Lambda_{ab}&=&\beta^{-2}r^2\nabla_a\nabla_bY_{lm}+f_{ab}\beta^{-1}\label{corro}
\end{eqnarray}
where in equation \ref{corro} we have corrected a typo of \citet{ush}.

From this expansion one can calculate the quadrupole moment $Q_{22}=\int\delta\rho r^4dr$ associated with the deformation
\begin{equation}
Q_{22}=\int\left\{\frac{r^4}{g(r)}\left[\frac{3}{2}\frac{dt_{rr}}{dr}-\frac{4}{\beta}\frac{d t_{r\perp}}{dr}-\frac{r}{\beta}\frac{d^2t_{r\perp}}{dr^2}+\frac{1}{3}\frac{dt_{\Lambda}}{dr}+t_{rr}\frac{3}{r}-\frac{\beta}{r}t_{r\perp}\right]+\frac{\delta\Phi}{g(r)}\frac{d\rho}{dr}\right\}dr
\end{equation}
In \citet{ush} the authors now make the Cowling approximation, thus choose to neglect the perturbations of the gravitational potential $\delta\Phi$, and integrate by parts by imposing that the shear modulus $\mu$ vanishes at the base of the crust. With this condition the surface terms vanish and the final expression for the maximum quadrupole of \citet{ush} is
\begin{equation}
Q_{22}=-\int_{rb}^{R}\frac{r^3}{g}\left[\frac{3}{2}(4-U)t_{rr}+\frac{1}{3}(6-U)t_{\Lambda}+\sqrt{\frac{3}{2}}\left(8-3U-\frac{1}{3}U^2-\frac{r}{3}\frac{dU}{dr}\right)t_{r\perp}\right]dr
\label{quad}\end{equation}
where $U=\frac{d\ln{g}}{d\ln{r}}+2$.
It is important to note that we have assumed that the surface terms in the integral vanish because we have assumed (as in \cite{ush}) that the shear stress vanishes as we approach the base of the crust, however this is not an accurate description of the physical situation. The shear modulus in fact increases as one approaches the base of the crust from the exterior regions of the star (as can be seen in figure \ref{mu}) and the transition to $\mu=0$ in the core may be very sharp; in this case the surface terms may be important.

\subsection{Maximum strain and the Von Mises criterion}

Up to now we have considered the purely elastic response of the crust. However real solids behave elastically only up to a maximum strain $\sigma_{\mbox{max}}$ after which they either crack or deform plastically. In order to estimate the maximum size of the mountain that can be built we shall assume that the crust cracks upon reaching a certain yield strain. Let us investigate this in detail, following once again \citet{ush}.  
We will consider the strain tensor rather than the stress tensor, as this allows us to use a simple criterion to establish when the crust breaks. The strain tensor is defined as:
\begin{equation}
\sigma_{ab}=t_{ab}/\mu
\label{strain}\end{equation}
where $\mu$ is the shear modulus. If we define $\bar{\sigma}$ as $\bar{\sigma}^2=\frac{1}{2}\sigma_{ab}\sigma^{ab}$, the Von Mises criterion (\cite{ush} and references therein) then states that the crust will yield when
\begin{equation}
\bar{\sigma}\ge \bar{\sigma}_{\mbox{max}}
\label{massi}
\end{equation}
Other criteria exist, such as the Tresca criterion which depends on the difference between the maximum and minimum eigenvalue of the strain tensor, but they are expected to lead to similar qualitative results (\cite{ush}).
If we assume the physical variables to be the real part of our complex variables we can write (\citet{ush})
\begin{eqnarray}
\sigma_{ab}\sigma^{ab}&=&\frac{3}{2}\sigma_{rr}^2[\Re(Y_{lm})]^2+\sigma_{r\perp}^2[\Re(f_{ab})]^2+\sigma_{\Lambda}^2[\Re(\Lambda_{ab}+\frac{1}{2}Y_{lm}e_{ab})]^2
\label{i2}\end{eqnarray}
where $\sigma_{rr}$, $\sigma_{r\perp}$ and $\sigma_{\Lambda}$ are the components of the strain tensor along the tensor spherical harmonics defined in equation (\ref{spando}).
By then assuming that the quadrupole is maximised when the equality of equation (\ref{massi}) is satisfied, i.e. when $\bar{\sigma}=\bar{\sigma}_{\mbox{max}}$, \citet{ush} find that the maximum is obtained when:
\begin{eqnarray}
\sigma_{rr}&=&\sqrt{\frac{32\pi}{15}}\bar{\sigma}_{\max}\\
\sigma_{r\perp}&=&\sqrt{\frac{16\pi}{5}}\bar{\sigma}_{\max}\\
\sigma_{\Lambda}&=&\sqrt{\frac{96\pi}{5}}\bar{\sigma}_{\max}
.\label{maxsol}\end{eqnarray}
The Von Mises criterion, together with the expression for the maximum quadrupole in (\ref{quad}), thus allowed \citet{ush} to conclude that the maximum quadrupole is \begin{equation}
Q_{22}^{\mbox{max}}\approx 10^{38} \mbox{g cm}^2 (\sigma_{\mbox{max}}/10^{-2})
\end{equation}
no matter how the strain arises and under the assumption that all the strain is in the $Y_{22}$ spherical harmonic; strain in other harmonics would push the crust closer to the yield point without contributing to the quadrupole. This estimate is thus an upper limit for the quadrupole the crust could sustain, and on the energy the star would emit in Gravitational Waves, which is related to the quadrupole by:
\begin{equation}
\dot{E}_{gw}=\frac{256\pi}{75}\frac{G\Omega^6}{C^5}Q_{22}^2
\end{equation}
where $\Omega$ is the rotation rate of the star.
The value $\sigma_{\mbox{max}}=10^{-2}$ was chosen because it puts $Q_{22}$ in the right range to balance the accretion torque in the LMXBs, the quadrupole required to do this is in fact (\citet{ush})
\begin{equation}
Q_{gw}=1.6\times 10^{38} \mbox{g cm$^2$} (300 \mbox{Hz}/\nu_{spin})^{5/2}
\end{equation}
if we assume the most basic accretion torque model, as presented for example in \citet{us}.
This would mean allowing the spin equilibrium periods of the LMXBs to be set by gravitational waves, and would explain the narrow distribution of such spin periods. It is, however, also possible to explain the spin distribution by using a more refined accretion torque model, that requires no gravitational waves (\citet{us}).
One should keep in mind that $\sigma_{\mbox{max}}$ is a very poorly constrained parameter, expected to be in the range $\sigma_{\mbox{max}}=10^{-5}-10^{-2}$ (\citet{numeri}).

\section{Accreted vs. Non-accreted crust}

\begin{figure}
\centerline{\includegraphics[height=7cm,clip]{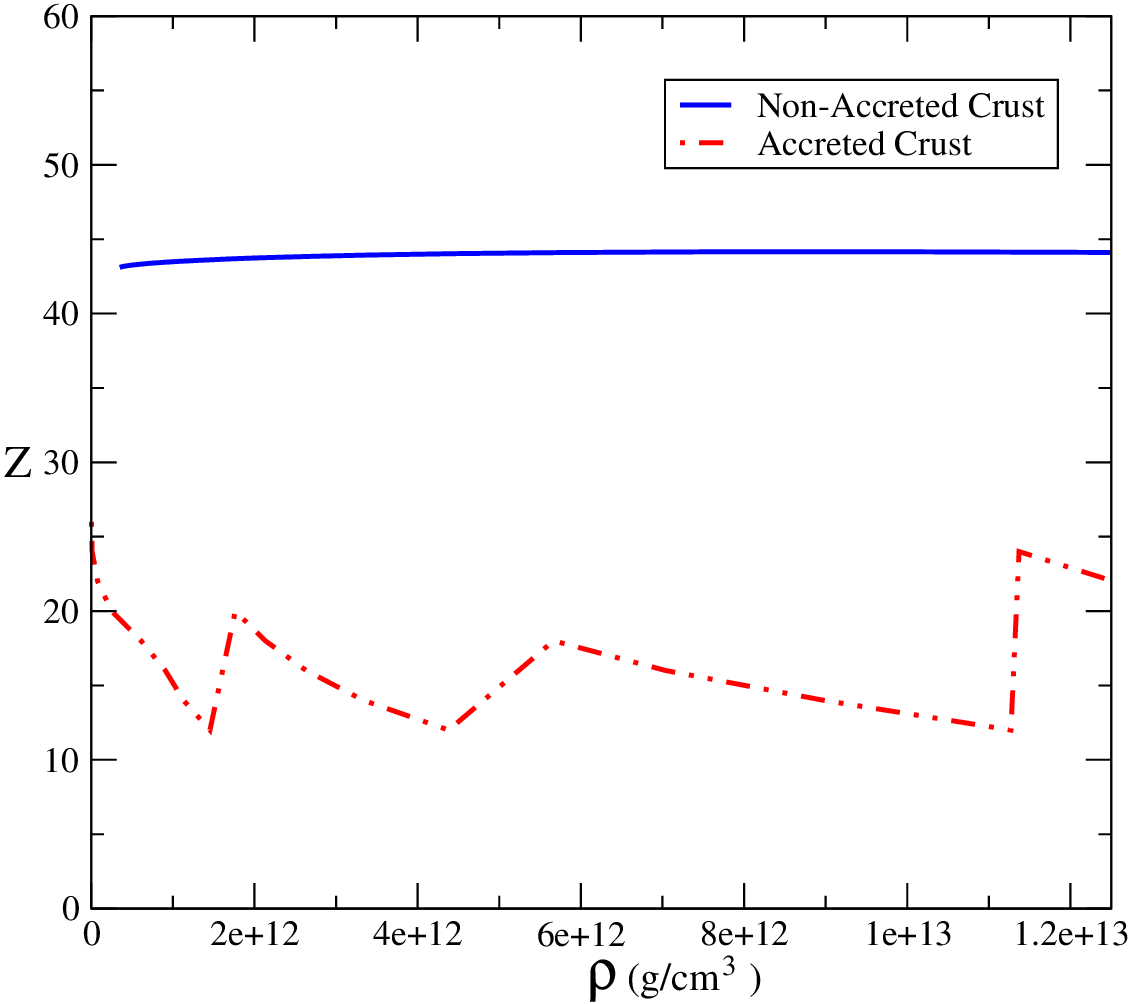}
\includegraphics[height=7cm,clip]{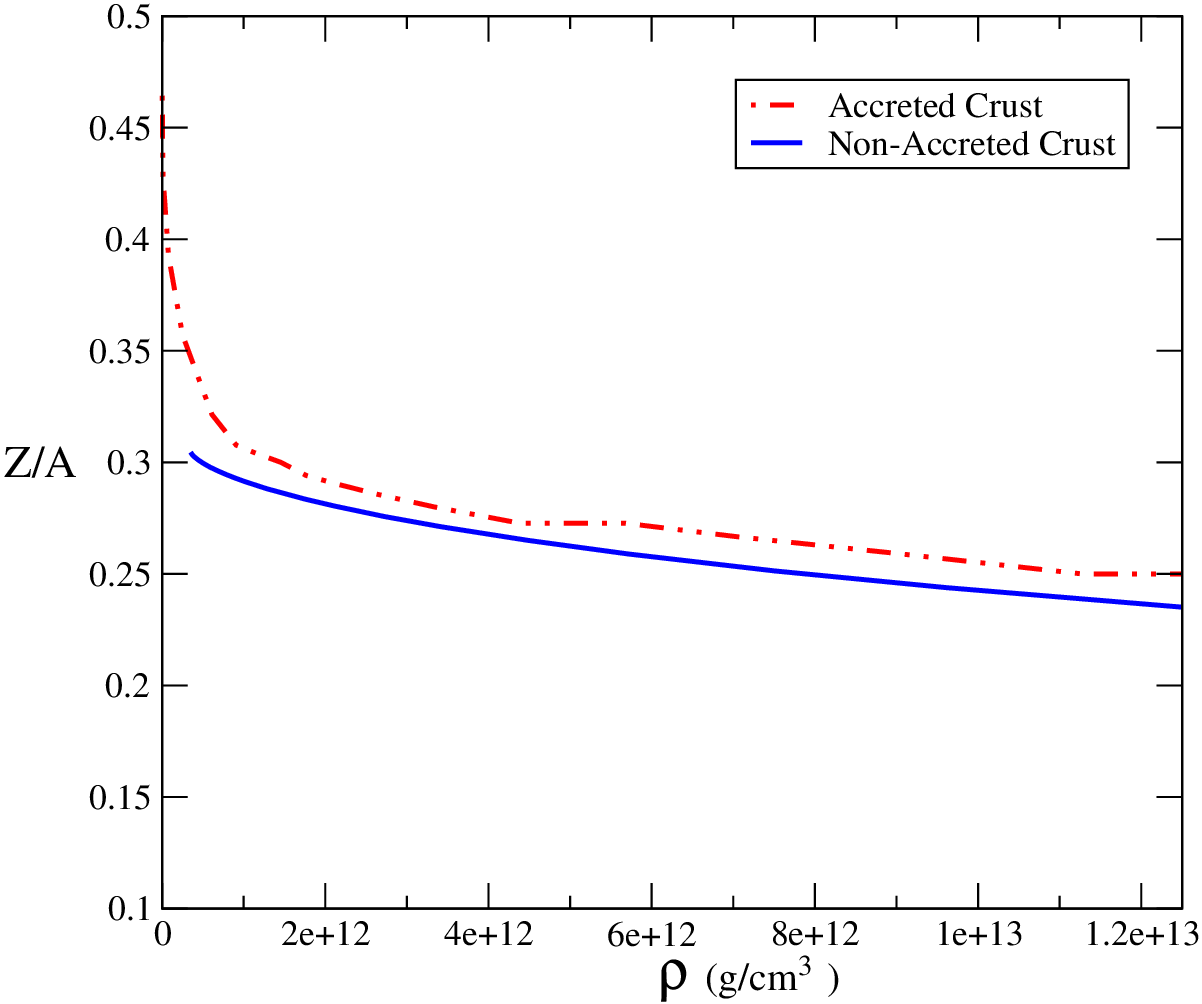} }
\caption{Comparison of the proton number Z and of the ratio Z/A for an accreted and a non-accreted crust. EOS from \citet{Duc} for the non-accreted case and from \citet{acro} for the accreted case.} 
\label{crust}
\end{figure}

\begin{figure} 
\centerline{\includegraphics[height=7cm,clip]{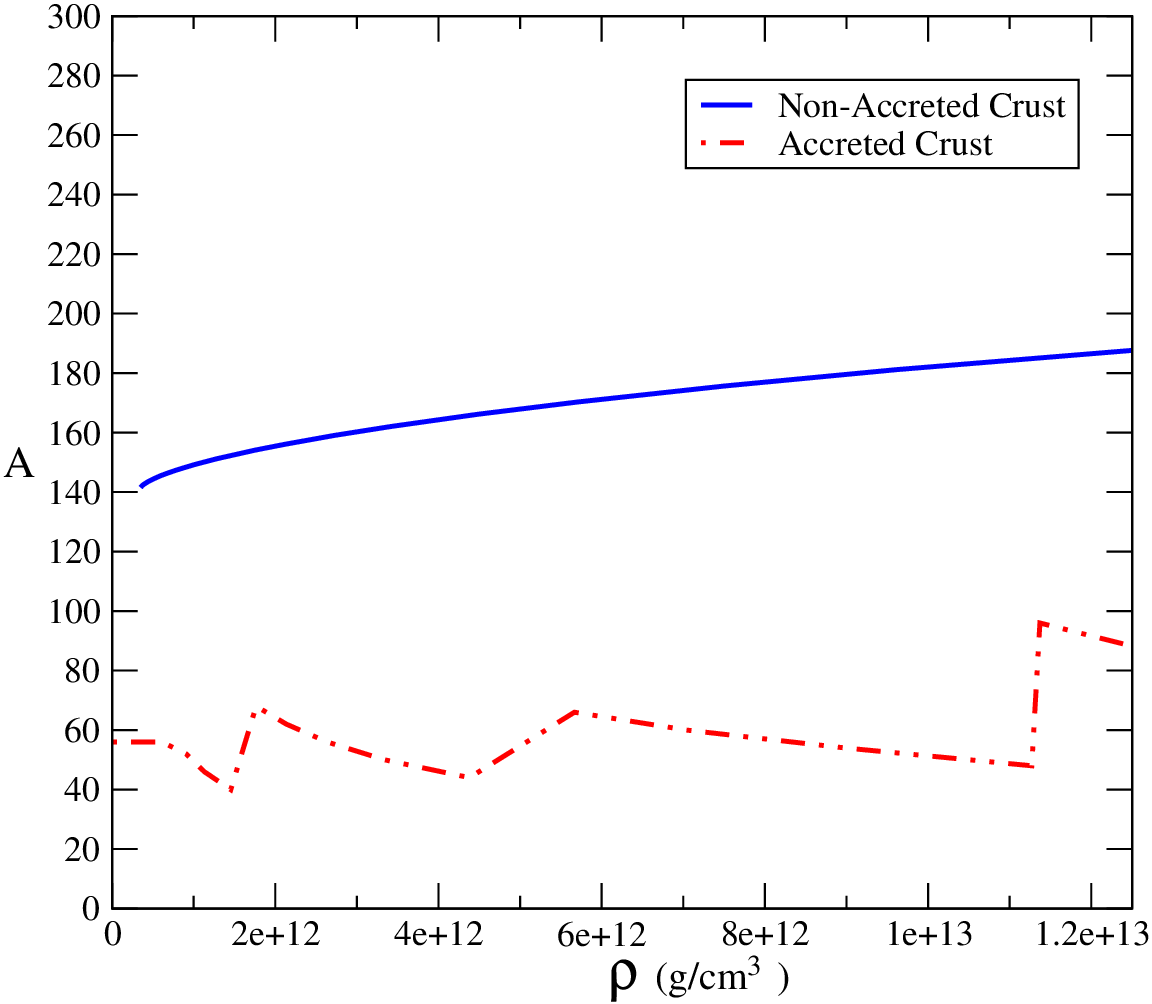}
\includegraphics[height=7cm,clip]{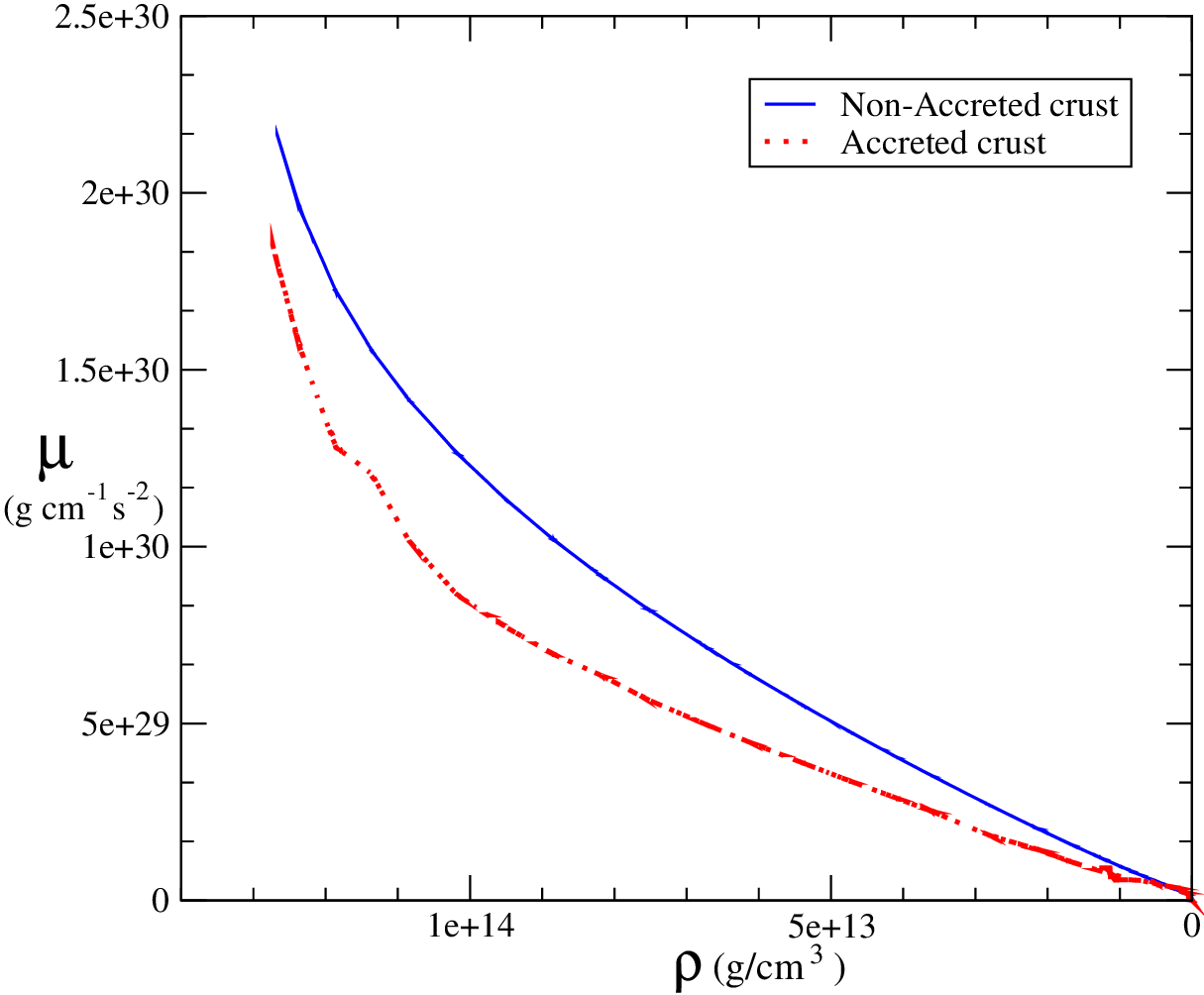} }
\caption{Comparison of the atomic number A and the shear modulus $\mu$ for an accreted and a non-accreted crust. The EOS is taken from \citet{Duc} for the non-accreted case and from \citet{acro} for the accreted case, and then extrapolated as described in the text.}
\label{mu}
\end{figure}

Having reviewed the formalism of \citet{ush}, let us apply it to the problem of determining the maximum quadrupole that an accreted and a non-accreted crust can sustain, and present some of the issues involved.
The maximum quadrupole defined in equation (\ref{quad}) depends on an integral over the crust involving the quantities $g$ and $\mu$ (which is contained in the components of the stress tensor $t_{ab}=\mu \sigma_{ab}$). These quantities depend on the equation of state and composition of the crust, which are quite different in the accreted and non-accreted cases (\citet{Haensel}). The equation of state, for example, determines the thickness of the crust, while the composition of the crust determines the shear modulus (\citet{ogata}):
\begin{equation}
\mu=0.1194\left(\frac{3}{4\pi}\right)^{1/3}\left(\frac{1-X_n}{A}n_b\right)^{4/3}(Ze)^2
\end{equation}
where $X_n$ is the fraction of neutrons outside nuclei, $n_b$ is the baryon density, $Z$ is the proton number and $A$ is the atomic number. These quantities are plotted in figures \ref{crust} and \ref{mu}.
One could then ask which kind of crust could support a larger deformation, which is the same as asking which kind of neutron star is potentially a ``better'' source of gravitational waves. It has been suggested that an accreted crust would sustain a smaller ``mountain'' as $Z$ is much lower than in a non-accreted crust (\citet{sato}).
To answer the question we can evaluate the expression in equation (\ref{quad}) in the two cases, using the results of \citet{Duc} and \citet{acro} for the composition and EOS of the crust. The main obstacle is that the composition of the accreted crust is only given up to neutron drip density, because at higher densities the equation of state is basically that of a neutron gas, and thus can be matched to that of a non-accreted crust. However, even if the EOS is the same, the composition (specifically $Z$ and $A$) is not (\citet{acro}). In order to obtain an estimate we have extrapolated the results available up to neutron drip density to higher densities by noting that the ratio $Z/A$ is roughly constant in the two kinds of crust and assuming that it will remain so even in the denser regions at the base of the crust. Having done this we can calculate the maximum quadrupole for the two cases. To do this we need to integrate the equations of hydrostatic equilibrium for the star (as the core is unperturbed) and we can use Newtonian equations or the TOV equations. We shall examine both cases in order to obtain an estimate of how important relativistic effects can be. The question is relevant, because even though relativistic effects are not very important in the crust, they do contribute significantly to the structure of the star, and thus affect parameters such as the crust thickness which are relevant for our calculation. It would then not seem appropriate to use realistic equations of state if we are still describing the core with Newtonian equations of equilibrium, as the relativistic effects we are neglecting may be as large as the corrections we expect from the equation of state. We will always use Newtonian equations in the crust.  

\begin{table}
\begin{tabular} {l |l |l |l}
\hline
&Accreted crust&Non Accreted crust & Accreted crust\\
\hline
TOV&&&\\
\hline
Mass & $1.4 M_{\odot}$& $1.4 M_{\odot}$& $1.6 M_{\odot}$\\
Radius &$12.56$ km&$12.3$ km&$12.3$ km\\
Crust Thickness&$1.76$ km&$1.5$ km&$1.5$ km\\
Q$_{\mbox{max}}$ &$1.6\times 10^{38}\left(\frac{\sigma}{10^{-2}}\right){\mbox{g cm}^2}$&$2.0\times 10^{38}\left(\frac{\sigma}{10^{-2}}\right){\mbox{g cm}^2}$&$1.1\times 10^{38}\left(\frac{\sigma}{10^{-2}}\right){\mbox{g cm}^2}$\\
$\epsilon$&1.1$\times 10^{-7}$&1.5$\times 10^{-7}$&7.3$\times 10^{-8}$\\
\hline
NEWTONIAN&&&\\
\hline
Mass & $1.4 M_{\odot}$& $1.4 M_{\odot}$& $1.7 M_{\odot}$\\
Radius &$14.1$ km&$14$ km&$14.3$ km\\
Crust Thickness&$2.2$ km&$1.9$ km&$1.9$ km\\
Q$_{\mbox{max}}$ &$2.8\times 10^{38}\left(\frac{\sigma}{10^{-2}}\right){\mbox{g cm}^2}$&$4.2\times 10^{38}\left(\frac{\sigma}{10^{-2}}\right){\mbox{g cm}^2}$&$2.5\times 10^{38}\left(\frac{\sigma}{10^{-2}}\right){\mbox{g cm}^2}$\\
$\epsilon$&1.6$\times 10^{-7}$&2.4$\times 10^{-7}$&1.1$\times 10^{-7}$\\
\hline
\end{tabular}\caption{The first two columns show maximum quadrupole for two stars of equal mass, one with an accreted crust and one with a non-accreted crust. The last column is a star with an accreted crust of the same thickness as that of the star with a  non-accreted crust, but a different mass. Both Newtonian equations and the TOV equations are used for the core, in order to illustrate GR effects. The equation of state is taken from \citet{Duc} for the core and for the non-accreted crust, while the equation of state for the accreted crust is taken from \citet{acro}. $\epsilon$ is the ellipticity $\frac{I_x-I_y}{I_0}$.}\label{comparo1} 

\end{table}
As we can see from Table \ref{comparo1} the non-accreted crust allows for a slightly larger quadrupole, but the results are essentially the same for the two models. This is mainly due to the fact that the slight decrease in $\mu$ (as seen in figure \ref{crust}) for the accreted crust is compensated for by it being thicker.We can try to compare the case of two stars with a crust of the same thickness and, as is obvious from Table \ref{comparo1}, in this case there is approximately a factor 2 difference. This is close to what would be expected, as \citet{sato} has predicted that the shear modulus of an accreted crust would be approximately half that of a non accreted crust, even if in reality this effect is counterbalanced by the accreted crust being thicker. However it is important to note that the result depends on how we extrapolate the composition at the base of the accreted crust. In order to confirm this result it would therefore be necessary to have a detailed calculation of the composition of the crust in this region.
From Table \ref{comparo1}, it is also clear that relativistic effects can play a role in this calculation, as they appear to make a difference of factors of up to a few in the final result (mainly because the crust is thicker in the Newtonian case, and the radius of the star is larger).
\section{Validity of the Cowling approximation and surface terms}

Following \citet{ush}, in the previous calculation we adopted the Cowling approximation. Such an approximation is often used when discussing oscillations, in cases where one expects only the surface layers to take part in the motion. Then the high density interior remains essentially unperturbed and one can neglect the effect of perturbations to the gravitational potential compared to other forces. We could expect this approximation to be valid also in our (non-oscillatory) context. This simplifies the calculation considerably, because it means that the liquid core of the star cannot support any kind of perturbation (with no shear and no $\delta\Phi$ there would be no restoring force for a pressure perturbation $\delta p$). This allows us to consider only the crust. In \cite{ush} the authors consider the effect of the approximation and conclude that retaining the perturbations of the gravitational potential could lead to changes in the quadrupole from $20\%$ to $200\%$. A calculation of the strain that arises in the crust, dropping the Cowling approximation, is carried out by \citet{link}, but for the case of a perturbation with a $Y_{20}$ angular dependance, given by rotation. In order to investigate the effects of such an approximation for our problem we will present a series of model calculations where the $\delta\Phi$ terms have been retained. These simple examples can be solved analytically and allow us to estimate how much the Cowling approximation can affect the final solution.

\subsection{Constant Density and Shear}

Let us first of all consider a constant density, elastic sphere. In this case we will have \begin{eqnarray}
&&\rho(r)=\rho\Theta(R-r)\;\;\; \mbox{and}\;\;\; \mu(r)=\mu\Theta(R-r)\\
&&{\mu}=\mbox{Const}
\end{eqnarray}
where $\Theta(R-r)$ is the Heaviside step function and we have taken the shear modulus constant over the whole star rather than just the crust, in order to make the algebra simpler. $R$ is the stellar radius.
In addition to equation (\ref{elas}) we now need to solve the perturbed Poisson equation
\begin{equation}
\nabla^2\delta\Phi=4\pi G\delta\rho 
\end{equation}
Let us examine the source term containing $\delta\rho$. Take for the components of the strain tensor the maximum from (\ref{maxsol}). This means that the stress tensor is simply $t_{ab}=\mu\sigma_{ab}$ and depends on $r$ only through $\mu$. If we assume that the crust is everywhere strained to the maximum and insert our definitions for $\mu$ and $\rho$ into equation (\ref{elas}) we obtain
\begin{eqnarray}
\delta\rho&=&-\frac{\delta(R-r)\mu}{g(r)}\left[\frac{3}{2}\sqrt{\frac{32\pi}{15}}\bar{\sigma}_{\max}-\frac{4}{\beta}\sqrt{\frac{16\pi}{5}}\bar{\sigma}_{\max}+\frac{1}{3}\sqrt{\frac{96\pi}{5}}\bar{\sigma}_{\max}\right]+\nonumber\\
&&+\frac{\Theta(R-r)}{g(r)}\left[\frac{3}{r}\sqrt{\frac{32\pi}{15}}\bar{\sigma}_{\max}-\frac{\beta}{r}\sqrt{\frac{16\pi}{5}}\bar{\sigma}_{\max}\right]\nonumber\\&&-\frac{\delta^{'}(R-r)}{g(r)}\frac{r}{\beta}\sqrt{\frac{16\pi}{5}}\bar{\sigma}_{\max}-\frac{\delta(R-r)}{g(r)}\rho\delta\Phi
\label{complete}\end{eqnarray}


Returning to Poisson's equation, in the $l=m=2$ case we have to solve:
\begin{equation}
\delta\Phi^{''}+\frac{2}{r}\delta\Phi^{'}-\frac{6}{r}\delta\Phi=-\delta(R-r)\left[\frac{3\mu\bar{\sigma}_{max}}{r\rho}\sqrt{\frac{8\pi}{15}}+\frac{3}{r}\delta\Phi\right],
\end{equation}
subject to the conditions that the solution should be regular at the centre of the star, fall off at infinity and be continuous at the surface of the star. The step in the first derivative of $\delta\Phi$ can be computed by integrating between $R-\epsilon$ and $R+\epsilon$ and then taking the limit $\epsilon\longrightarrow 0$; this gives
\begin{equation}
\delta\Phi^{'}(R+\epsilon)-\delta\Phi^{'}(R-\epsilon)=-\left[\frac{3\mu\bar{\sigma}_{max}}{R\rho}\sqrt{\frac{8\pi}{15}}+\frac{3}{R}\delta\Phi(R)\right]
\end{equation}
The solutions of the homogeneous equation are of the form $Ar^2+\frac{B}{r^3}$
We then clearly have $Ar^2$ for the interior and $B/r^3$ for the exterior.
Matching at the surface gives us $B=AR^5$ which can be used to compute the step in the first derivative and obtain the interior solution
\begin{equation}
\delta\Phi=\frac{3}{2}\frac{\mu\bar{\sigma}_{max}}{R^2\rho}\sqrt{\frac{8\pi}{15}}r^2
\end{equation}
We can now substitute this back into equation (\ref{complete}) and calculate:

\begin{equation}
Q_{22}=\left[\frac{9\mu\bar{\sigma}_{max}R^3}{4\pi G\rho}\sqrt{\frac{8\pi}{15}}-\frac{3}{4\pi G}\frac{3}{2}\frac{\mu\bar{\sigma}_{max}R^3}{\rho}\sqrt{\frac{8\pi}{15}}\right]\equiv Q_{sh}+Q_{\Phi}\label{qphi}
\end{equation}
which gives us the ratio
\begin{equation}
\frac{Q_{\Phi}}{Q_{sh}}=-\frac{1}{2}
\end{equation}
where $Q_{sh}$ is the part of the quadrupole we would get in the Cowling approximation and $Q_{\Phi}$ is the part that comes from the terms involving $\delta\Phi$, as defined in equation (\ref{qphi}).
It would thus appear that, in this case, by making the Cowling approximation we could overestimate $Q$ by a factor of 2.

Let us see what this means in terms of the ellipticity $\epsilon=\frac{I_x-I_y}{I_0}$. We can write the torque acting on a constant density ellipsoid as
\begin{equation}
\dot{J}=\frac{32G}{5c^5}\epsilon^2({2/5MR^2})^2\Omega^5
\end{equation}
However we can also write this as 
\begin{equation}
\dot{J}=\frac{256\pi}{75}{G\Omega^5Q_{22}^5}{c^5}
\end{equation}
If we compare these two expressions we see that, in the constant density case
\begin{eqnarray}
\epsilon&=&\sqrt{30\pi}\frac{Q_{22}}{3R^2M}\\
&=&\frac{9\mu\bar{\sigma}_{max}}{4G\pi\rho^2 R^2}\left(1+\frac{Q_\Phi}{Q_{sh}}\right)\label{epso}
\end{eqnarray}
Let us take as canonical values $\mu=10^{30}$ $dyne/cm^2$, $\rho=10^{15}$ $g/cm^3$, $R=10$ $km$. Substituting these into the above expression gives 
\begin{equation}
\epsilon=5.3\times10^{-8}\left(\frac{\bar{\sigma}_{max}}{10^{-2}}\right)
\end{equation}

\subsection{Constant Shear in the Crust}
A slight improvement one can make is to impose that the shear modulus vanishes outside the crust. We will still take a constant density but we shall take $\mu$ of the form:
\begin{equation}
\mu(r)=\mu\left[\Theta(R-r)-\Theta(r_b-r)\right]
\end{equation}
The calculation is similar to the previous one, but we now have to match the solutions at $R$ and $r_b$ and impose the step in the derivative at these two points.
It is in fact important to note that as we now have a step at the base of the crust, this will lead to surface terms when we take derivatives of the shear modulus. These terms are important, as they are not present in the calculation of \citet{ush}, where $\mu \longrightarrow 0$ at the base of the crust.
From this calculation we obtain
\begin{eqnarray}
Q_{sh}&=&\frac{3\bar{\sigma}_{max}\mu (R^3-rb^3)}{10G\rho}\sqrt{\frac{30}{\pi}}\\
Q_{\Phi}&=&-\frac{3\bar{\sigma}_{max}\mu (R^3-r_b^3)}{20G\rho}\sqrt{\frac{30}{\pi}}
\end{eqnarray}
Again, the ratio between the two is
\begin{equation}
\frac{Q_{\Phi}}{Q_{sh}}=-\frac{1}{2}
\end{equation}
and
\begin{equation}
Q_{sh}=1.9\times 10^{37} \mbox{g cm}^{2}
\end{equation}
In terms of $\epsilon$ we have
\begin{equation}
\epsilon=\frac{\sqrt{30}Q}{4\pi^{1/2}R^5\rho}=\frac{\sqrt{30}Q_{sh}}{4\pi^{1/2}R^5\rho}\left(1+\frac{Q_\Phi}{Q_{sh}}\right)
\end{equation}
which gives us, for the canonical values we have chosen and $r_b=1$ $km$:
\begin{equation}
\epsilon=7.2\times10^{-9}\left(\frac{\bar{\sigma}_{max}}{10^{-2}}\right)
\end{equation}
which is nearly an order of magnitude smaller than the previous result for the elastic sphere. So the shell is, as is intuitive, ``weaker'' than the sphere.

\subsection{n=1 Polytrope}
A slightly more realistic model is that given by an $n=1$ polytrope
\begin{equation}
p=K\rho^2
\end{equation}
 which leads to the density profile:
\begin{equation}
\rho(r)=\rho_c\frac{\sin{(\pi r/R)}}{r\pi}
\end{equation}
Let us still take $\mu(r)=\mu\left[\Theta(R-r)-\Theta(r_b-r)\right]$ with $\mu$ constant.
The homogeneous Poisson equation thus becomes, in the interior:
\begin{equation}
\delta\Phi^{''}+\frac{2}{r}\delta\Phi^{'}-\frac{6}{r^2}\delta\Phi+\frac{\pi^2}{R^2}\delta\Phi=0
\end{equation}
which admits solutions of the form
\begin{eqnarray}
\delta\Phi&=&\frac{A}{r^3}\left[(3R^2-\pi^2r^2)\cos{(\pi r/R)}+3\pi Rr\sin{(\pi r/R)}\right]+\\
&&+\frac{B}{r^3}\left[(3R^2-\pi^2r^2)\sin{(\pi r/R)}-3\pi Rr\cos{(\pi r/R)}\right]
\end{eqnarray}
By expanding this solution with a power series around the origin we can see that the regular part is that involving the constant $B$ (thus $A=0$). This must be matched with the whole solution at $r_b$, and we must then match with the exterior solution $C/r^3$ at the stellar surface $R$. Finally we must impose the steps in the derivative at $r_b$ and $R$.
The resulting solution is quite complicated, but leads to
\begin{equation}
\frac{Q_\Phi}{Q_{sh}}\approx 0.1
\end{equation}
for a radius $R$=10 km and a crust thickness of 1 km. Examining the ratio $\frac{Q_\Phi}{Q_{sh}}$ for different values of the crust thickness leads to very similar results and allows us to conclude that  including $\delta\Phi$ always slightly increases the value of the quadrupole.
If we now calculate an $\epsilon$ we obtain:
\begin{equation}
\epsilon=\frac{\sqrt{30}Q}{4\pi^{1/2}R^5\rho}=\frac{\sqrt{30}Q_{sh}}{4\pi^{1/2}R^5\rho}\left(1+\frac{Q_\Phi}{Q_{sh}}\right)
\end{equation}
which gives us, with typical parameters for the crust and $r_b=1$ $km$, 
\begin{equation}
\epsilon=3.1 \times 10^{-7}\left(\frac{\bar{\sigma}_{max}}{10^{-2}}\right)
\end{equation}
which is an order of magnitude larger than in the constant density cases.

\subsection{Density dependent Shear}

\begin{figure} 
{\includegraphics[height=6cm,clip]{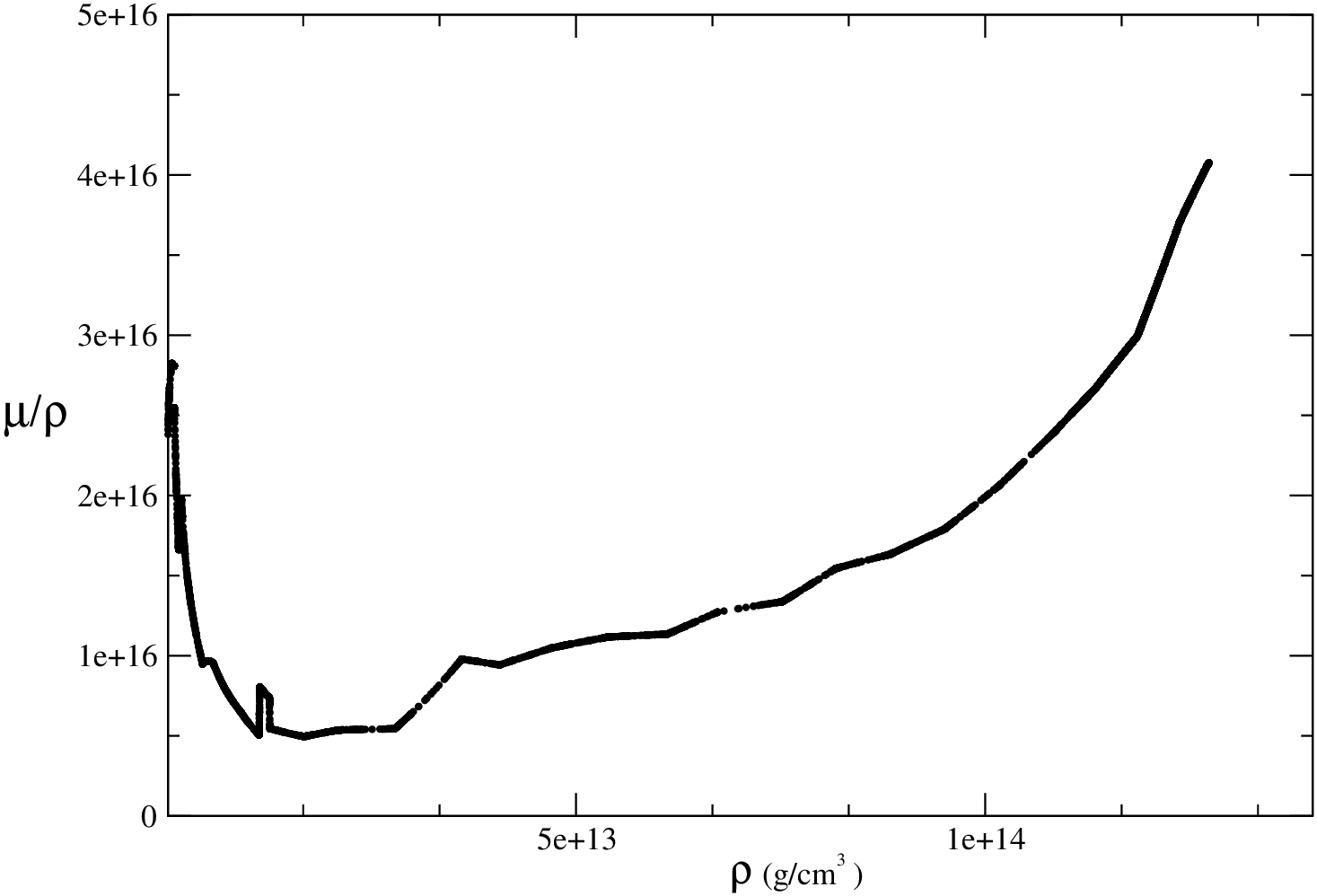} }
\caption{ The ratio $\mu/\rho$, which is taken to be constant in our approximation, is plotted for the case of a non-accreted crust (the case of an accreted crust is qualitatively similar). As we can see it is approximately constant to within factors of two.} 
\label{musurho}
\end{figure}

Let us finally take a slightly more realistic model with a density dependent shear. We will take $\mu/\rho$ to be constant which, as shown in Figure \ref{musurho}, is not a bad approximation. This gives us
\begin{equation}
\mu=C\rho\Theta(r-r_b)
\end{equation}
where the step function ensures that the shear modulus vanishes in the core ($r_b$ is the radius at the base of the crust). There is no need for a step at the surface, as the density goes to zero there anyway. We will take $C=10^{16}$ $cm^2/s^2$.
As in the previous section we will take the equation of state to be an $n=1$ polytrope, thus obtaining:
\begin{equation}
\mu=C\rho_c\frac{\sin{(\pi r/R)}}{r\pi}\Theta(r-r_b)
\end{equation}
The components of the stress tensor will now also depend on $r$ through the shear modulus.
\begin{eqnarray}
t_{rr}&=&\frac{4}{15}\Theta(r-r_b)\sigma\frac{C\rho_c\sin{(\pi r/R)}\sqrt{30}R}{r\sqrt{\pi}}\\
t_{r\perp}&=&\frac{4}{5}\Theta(r-r_b)\sigma\frac{C\rho_c\sin{(\pi r/R)}\sqrt{5}R}{r\sqrt{\pi}}\\
t_{\lambda}&=&\frac{4}{5}\Theta(r-r_b)\sigma\frac{C\rho_c\sin{(\pi r/R)}\sqrt{30}R}{r\sqrt{\pi}}
\end{eqnarray}
The procedure we have to follow is identical to that of the previous sections, we have to solve Poisson's equation in the core, in the crust and outside the star. We must then impose the boundary conditions at the edge of these regions. These are the same as in the preceding sections, i.e. we need to impose the continuity of $\delta\Phi$ and the step in the derivative at the base of the crust.

This allows us to compute the part of the quadrupole that comes from the perturbations of the gravitational potential $Q_{\Phi}$.
For our canonical values we have
\begin{equation}
Q_{\Phi}=-2.3\times 10^{38}\;\;\;\mbox{g cm$^2$}
\end{equation}
We must also calculate the quadrupole that comes from the shear terms alone. As we have mentioned before this contribution involves derivatives of the stress tensor, it also involves delta functions and their derivatives. This means that we have two contributions: a surface term that comes from such delta functions and a term that comes from the interior integration. Let us examine this case in more detail, as it is a good example of how important the surface terms can be.

For the stellar parameters given above we find that the integral term is:
\begin{equation}
Q_{int}=-3.3\times 10^{38}\;\;\;\mbox{g cm$^2$}
\end{equation}
while the surface term has opposite sign and gives
\begin{equation}
Q_{\Sigma}=4 \times 10^{38}\;\;\;\mbox{g cm$^2$}
\end{equation}
The two thus add up to 
\begin{equation}
Q_{sh} =7.4\times 10^{37}\;\;\;\mbox{g cm$^2$}
\end{equation}
the large cancellations due to the surface term lead to a total quadrupole where the dominant contribution is that given by the perturbations of the gravitational potential
\begin{eqnarray}
\frac{Q_{\Phi}}{Q_{sh}}&=&-3\\
Q_{tot}&=&-1.5\times 10^{38}\;\;\;\mbox{g cm$^{2}$}
\end{eqnarray}
corresponding to an $\epsilon$ of
\begin{equation}
\epsilon=\frac{\sqrt{30\pi}Q}{3 R^2 M}=6.3\times 10^{-8} 
\end{equation}
Which is smaller than the previous result and is a direct consequence of the cancellations that occur at the base of the crust. It is thus quite clear that the correct treatment of the boundary can have an important effect on this calculation. In particular it would appear that one cannot take $\mu\longrightarrow 0$ at the base of the crust, as the step is actually quite sharp (as shown in Figure \ref{mu}) and we have seen that this can lead to large surface terms.
We can, however, conclude, similarly to \citet{ush}, that including the perturbations of the gravitational potential $\delta\Phi$ will affect the results by a factor $0.5$-$3$. Thus the Cowling approximation is not consistent, as the corrections given by $\delta\phi$ are not negligible. It must, however, be noted  that when we drop the Cowling approximation we are no longer guaranteed that the components of the strain tensor calculated in (\ref{elas}) will give us the maximum quadrupole.
\section{Perturbation formalism}

Up to now we have followed the method of \citet{ush} and simply taken the crust to be strained to the maximum throughout. However, as we have seen previously, the boundaries can have a strong impact on the result, so in order to carefully impose the boundary conditions at the surface of the star and at the base of the crust let us reformulate the problem in terms of perturbations in the crust.
We shall impose (in a similar way as was done to calculate rotational deformations in the slow-rotation approximation, as presented for example in \citet{tas}) that the star has a small quadrupolar deformation, such that we can define a new variable $a$ which is a function of the radial variable of the spherical background model
\begin{equation}
a(r)=r\left[1+\psi(r)Y_{22}(\theta,\phi)\right].
\end{equation}
We shall assume that $a(r)$ labels the equipotential surfaces of the gravitational potential (which will thus have the same deformed shape). It is important to note that in the core these surfaces also describe the ``isobaric'' and ``isopycnic'' surfaces. This is a consequence of the hydrostatic equations. In the crust, where we also have elastic terms, this is no longer true. In the crust, surfaces of constant $\Phi$ are NOT surfaces of constant density, so while $\epsilon$ is a good variable to use for the integration, it does not represent the ``shape'' of the star, which can be obtained by calculating the perturbations of the density ($\delta\rho$) or of the pressure (as we will assume the equation of state to be barotropic).
We expand the displacement vector along the orthonormal basis formed by the radial unit vector ${\hat{r}}$, $\frac{r}{\beta}{\nabla Y_{lm}}$ and $\frac{r}{\beta}{\hat{r}\times\nabla Y_{lm}}$. This allows us to use a flat background 3-metric $g_{ab}$.
We shall ignore the axial part, proportional to $\bar{r}\times\nabla Y_{lm}$ as this does not couple to the scalar perturbations such as $\delta\rho$ which are proportional to $Y_{lm}$. The displacement vector thus takes the form:

\begin{equation}
\xi^a=\xi_r(r)Y_{lm}r^a+\xi_{\perp}\nabla^aY_{lm}
\end{equation}

If we assume that the crust responds elastically it is useful to define the stress-energy tensor of the solid $\tau_{ik}$ by
\begin{equation}
\tau_{ik}=-pg_{ab}+t_{ab}
\end{equation}
where we have defined the trace-free tensor 
\begin{equation}
t_{ab}=\mu\left(\nabla_a\xi_b+\nabla_b\xi_a-\frac{2}{3}g_{ab}\nabla^c\xi_c\right)
\end{equation}
which describes the shear stress in the solid.
This allows us to write the equations of static balance as
\begin{equation}
\nabla^at_{ab}=\rho\nabla_b\Phi+\nabla_bp
\end{equation}
which we can also write in the form
\begin{equation}
\nabla^at_{ab}=\rho\nabla_b(\Phi+H)
\end{equation}
where we have introduced the enthalpy 
\begin{equation}
H=\int \frac{dp}{\rho(p)}.
\end{equation}
We also have to supply a barotropic equation of state
\begin{equation}
p=p(\rho)
\end{equation}
The remaining equations we need are the continuity equation and Poisson's equation
\begin{eqnarray}\delta\rho&=&-\partial_r\rho\xi_r-\rho\partial_r\xi_r+\rho\frac{\beta}{r}\xi_{\perp}-2\frac{\rho}{r}\xi_r\\
\nabla^2 \Phi&=&4\pi G\rho
\end{eqnarray}
and the condition that the unit vector $a^b$ given by the new radial variable is orthogonal to the equipotential surfaces of $\Phi$, or parallel the gradient of $\Phi$, i.e.
\begin{equation}
\nabla a \times \nabla \Phi=0
\label{level}\end{equation}
It is important to note again that $a$ is not related to the 'shape' of the star (as the deformations of the density are) because we now have elastic terms in our equations of force balance. However the introduction of the variable $\psi$ makes the equations for $\delta\Phi$ more tractable numerically.
We will now carry out first order perturbations, treating $\xi_a$ as a first order quantity. This corresponds to assuming that the background spherical configuration is at zero strain. We could imagine this to be plausible for a star which has been spun down by dipole radiation and been in an essentially non rotating configuration long enough for plastic flow to relax the crust. If this star then starts accreting it would be spun up and correspond to the situation we are describing. It is of course not relevant for the case of a spinning down star with a crust that formed while it was rapidly rotating and for which the reference zero-strain configuration would not be spherical. This problem is, however, a much more complicated one, as it would require us to know the reference shape of the star, which will in general depend on the history of the crust.

We shall assume that every perturbation has a $Y_{22}$ angular dependance, i.e. is of the form $\delta\rho(r,\theta,\phi)=\delta\rho(r)  Y_{22}(\theta,\phi)$. This is because we are interested in the quadrupolar part of $\delta\rho$ which will give the dominant contribution for gravitational waves. However it is worthwhile noting that, even though they do not contribute to the gravitational wave emission, perturbations with a different angular dependance (e.g. a $Y_{20}$ deformation due to rotation) still build up stress in the crust, and bring it closer to the breaking point, thus allowing a smaller quadrupolar deformation. For example, there could be a strong $Y_{20}$ deformation, as the star is rotating, and this could build up high levels of strain in the crust. It is not however easy to evaluate this quantitatively, as the exact level of strain depends on the reference shape of the crust (the unstrained configuration), which in turn depends on the history of the star.
Let us then write the perturbation equations we need ($\beta=\sqrt{l(l+1)}=\sqrt{6}$):

\begin{tabular}{l c l}
& & \\
$\nabla_bt^{ab}=\rho\nabla^a\left[(\delta H+\delta\Phi)Y_{22}\right]$ & &{Hydrostatic equilibrium}\\
& &\\
$\delta\rho=-\partial_r\rho\xi_r-\rho\partial_r\xi_r+\rho\frac{\beta}{r}\xi_{\perp}-2\frac{\rho}{r}\xi_r$& & Continuity\\
& &\\
$\delta p=\frac{dp(\rho)}{d\rho}\delta\rho=c_s^2\delta\rho$ & & Equation of state\\
& & \\
$\delta H=\frac{\delta p}{\rho}$& & {Enthalpy}\\
& & \\
$\delta \Phi=-\psi r\partial_r \Phi$ & & {Level surfaces (from eq.(\ref{level}))}\\
& & \\
$4\pi G\delta\rho=\partial_r^2\delta\Phi+(2/r)\partial_r\delta\Phi-(\beta^2/r^2)\delta\Phi$ & & {Poisson}\\
& & \\
& &\\
\end{tabular}

This gives us 7 equations for our 7 variables $\xi_r,\xi_\perp,\psi,\delta\rho,\delta p,\delta H,\delta\Phi$. Some of these equations are simply algebraic relations defining some quantities (such as the enthalpy). We can thus re-write the equations as three second order equations containing the variables $\xi_r,\xi_\perp$ and $\psi$. Once we have solved for these variables we can compute $\delta\rho$, for example from the continuity equation.
Let us write the stress tensor $t_{ab}$ in terms of our displacements:
\begin{eqnarray}
t_{ab}&=&\mu\left[g_{ab}\left(-2/3\frac{d\xi_r}{dr}+2/3\frac{\xi_{\perp}\beta}{r}+2/3\frac{\xi_r}{r}\right)Y_{lm}+\hat{r}_a\hat{r}_b\left(2\frac{d\xi_r}{dr}-2\frac{\xi_r}{r}\right)Y_{lm}+\left(\nabla_a\nabla_bY_{lm}\right)\frac{r}{\beta}2\xi_\perp+\right.\\
&&\left.\left(\hat{r}_a\nabla_bY_{lm}+\hat{r}_b\nabla_aY_{lm}\right)\left(\xi_r+\frac{d\xi_\perp}{dr}\frac{r}{\beta}+\frac{\xi_\perp}{\beta}\right)\right]
\end{eqnarray}

Working out some of the algebra we obtain for $\nabla_bt^{ab}$, breaking it down in its two components along $\hat{r}^a$ and $\nabla^aY_{22}$:
\begin{eqnarray}
\hat{r}_a\nabla_bt^{ab}&=&\frac{Y_{lm}}{3r^2}\left[{4}\frac{d\mu}{dr}\left(\frac{d\xi_r}{dr}r^2+r\beta\xi_{\perp}/2-r\xi_r\right)+\mu\left(\frac{d^2\xi_r}{dr^2}r^2-\frac{d\xi_{\perp}}{dr}\beta r+7\xi_{\perp}\beta+8\frac{d\xi_r}{dr}r-8\xi_r-\xi_r\beta^2\right)\right]
\end{eqnarray}
and
\begin{eqnarray}
 \nabla_a Y_{22}\nabla_bt^{ab}&=&\frac{1}{\beta r}\left[\frac{d\mu}{dr}\left(r\beta\xi_r+r^2\frac{d\xi_\perp}{dr}+r\xi_\perp\right)+\mu\left(\frac{d\xi_r}{dr}r\beta/3-4/3\beta^2\xi_\perp+8/3\xi_r\beta+2\frac{d\xi_\perp}{dr}r+\frac{d^2\xi_\perp}{dr^2}r^2\right)\right]
\end{eqnarray}
The hydrostatic equilibrium equations then give
\begin{eqnarray}
\hat{r}_a\nabla_bt^{ab}&=&\rho\partial_r(\delta H+\delta\Phi)\\
\nabla_a Y_{22}\nabla_bt^{ab}&=&\rho(\delta H+\delta\Phi)
\end{eqnarray}
Note that the term involving $\delta\rho\partial_r(H+\Phi)$ is not present because the background quantity $\partial_r(H+\Phi)$ vanishes in equilibrium (as the background spherical configuration is a zero-strain one).
At this point we can write the equations as a set of four algebraic equations and three coupled second order equations for the displacements:
\begin{eqnarray}
&&{\frac{d^2\xi_r}{dr^2}}\left(3c_s^2\rho+4\mu\right)\rho r^2 =-\left(2\rho\frac{d\mu}{dr}r\beta+7\rho\mu\beta+3c_s^2\beta\rho^2-3\frac{d}{dr}c_s^2\right){\xi_\perp}-\\
&&-\left(-4\rho\frac{d\mu}{dr}r-8\rho\mu-3\rho\mu\beta^2+3c_s^2\rho\frac{d^2\rho}{dr^2}r^2-3c_s^2\left(\frac{d\rho}{dr}\right)^2r^2+6\frac{d}{dr}c_s^2r\rho^2-6c_s^2\rho^2\right){\xi_r}-\nonumber\\
&&-\left(8\rho\mu r+3c_s^2\rho\frac{d\rho}{dr}r^2+4\rho\frac{d\mu}{dr}r^2+3\frac{d}{dr}c_s^2r^2\rho^2\right){\frac{d\xi_r}{dr}}+\nonumber\\
&&+\left(\rho\mu\beta r+3c_s^2\beta\rho^2r\right){\frac{d\xi_\perp}{dr}}+\left(3r^3\frac{dp}{dr}\frac{d\rho}{dr}\rho\right){\frac{d\psi}{dr}}-\left(3r^3c_s^2\frac{d\rho}{dr}\rho+12r^3\rho^3\pi G\right){\psi}\nonumber
\end{eqnarray}
\begin{eqnarray}
&&{\frac{d^2\xi_\perp}{dr^2}}3\left(\mu r^2\right)=-\left(-4\mu\beta^2+3\frac{d\mu}{dr}-3c_s^2\beta^2\rho\right){\xi_\perp}-\left(3\frac{d\mu}{dr}r\beta+8\mu\beta+3\beta c_s^2\frac{d\rho}{dr}r+6c_s^2\beta\rho\right){\xi_r}-\\
&&-\left(\mu\beta r+3\beta c_s^2\rho r\right){\frac{d\xi_r}{dr}}-\left(3\frac{d\mu}{dr}r^2+6\mu r\right){\frac{d\xi_\perp}{dr}}+\left(3\beta c_s^2\frac{d\rho}{dr}r^2\right){\psi}\nonumber
\end{eqnarray}
\begin{eqnarray}
&&{\frac{d^2\psi}{dr^2}}r^2\left({c_s^2\frac{d\rho}{dr}}\right)=\left( 4\pi G\beta\rho^2\right){
\xi_\perp}-\left(4\pi G r\frac{d\rho}{dr}\rho-8\pi G\rho^2\right){\xi_r}+\left(4r^2 G\frac {d\rho}{dr}\rho- \beta^2c_s^2+ 8\pi G\rho^2 r\right){\psi}-\\
&&+\left(8\pi G \rho^2 r\right){\frac{d\psi}{dr}}-\left(4\pi G\rho^2 r\right){\frac{d\xi_r}{dr}}\nonumber
\end{eqnarray}

If we make the Cowling approximation these equations reduce to those used by \citet{mac}.
\subsection{Boundary conditions}

Let us now move on to the key issue that was ignored in our previous analysis. 
We need to prescribe boundary conditions at the surface and at the base of the crust. To do this in a consistent manner we should solve the perturbation problem in the core and then impose the continuity of the gravitational potential and the continuity of the perpendicular and radial components of the traction vector. We assume that the perpendicular tractions vanish outside the star and also in the liquid core, as it cannot support shearing. The core could in principle sustain pressure and density perturbations, and these will appear in the radial component of the traction. 
The definition of the traction vector is
\begin{equation}
t^a=T^{ab}\hat{a}_b
\end{equation}
where $T^{ab}$ is now the complete stress tensor, i.e.
\begin{equation}
T^{ab}=-\delta pg^{ab}+t^{ab}
\end{equation}
This gives us
\begin{equation}
t^a=\hat{r}^a\left[\mu\left(\frac{4}{3}\frac{d\xi_r}{dr}-\frac{4}{3}\frac{\xi_r}{r}+\frac{2}{3}\frac{\xi_\perp}{r}\beta\right)+c_s^2\left(\frac{d\rho}{dr}\xi_r+\rho\frac{d\xi_r}{dr}-\rho\beta\frac{\xi_\perp}{r}\right)\right]+\nabla^aY_{lm}\mu\left(\xi_r+\frac{d\xi_\perp}{dr}\frac{r}{\beta}+\frac{\xi_\perp}{\beta}\right)
\end{equation}
Imposing that the component of this along $\nabla^aY_{22}$ must vanish gives us the condition (at the base of the crust):
\\
\begin{equation}
\frac{d\xi_\perp}{dr}(r_b)=-\frac{\xi_r(r_b)\beta}{r_b}-\frac{\xi_\perp(r_b)}{r_b}
\end{equation}
We then assume that the radial component of the traction is continous, thus equal to $t^a_{\mbox{core}}$. This leads to
\begin{equation}
t^a_{\mbox{core}}=\mu(r_b)\left(\frac{4}{3}\frac{d\xi_r}{dr}(r_b)-\frac{4}{3}\frac{\xi_r(r_b)}{r_b}+\frac{2}{3}\frac{\xi_\perp(r_b)}{r_b}\beta\right)+c_s^2(r_b)\left(\frac{d\rho}{dr}(r_b)\xi_r(r_b)+\rho\frac{d\xi_r}{dr}(r_b)-\rho\beta\frac{\xi_\perp(r_b)}{r_b}\right)
\end{equation}

We shall consider the full perturbation problem in the following sections, but first, for simplicity, let us consider some simple boundary conditions: we shall assume that the fluid core is \textbf{unperturbed} and remains spherical. Such a condition is not physical, but will allow us to consider a full GR fluid core (as we only need the background model) and quantify how important relativistic effects can be. However, it is not a natural condition since it needs a force acting at the surface as well as some forces to be acting at the base of the crust to keep it spherical. This is why we cannot impose the traction conditions but will rather impose that the radial component of the displacement ($\xi_r$) vanishes (i.e. the core remains spherical and does not 'move' into the crust) and that the shape of the inner edge of the crust remains spherical, which implies that $\delta\rho$ must vanish (remember that the background is already spherical, so there must be no $Y_{22}$ perturbations if it is to remain spherical).
We also impose continuity of the gravitational potential, and thus impose $\epsilon(r_b)=0$ (which corresponds to $\delta \Phi$ vanishing at the base of the crust). Our boundary conditions are thus 
\begin{eqnarray}
\xi_r&=&0\\
\frac{d\xi_r}{dr}&=&\beta\frac{\xi_\perp}{r_b}-\xi_r\frac{d\ln\rho}{dr}-2\frac{\xi_r}{r_b}\\
\psi&=&0
\end{eqnarray}

The shape can in fact be described by a variable defined to label the equipotential surface of the density or pressure (very much like $\psi$). We can thus define
\begin{equation}
\varepsilon_s=-\frac{\delta p}{r \partial_r p}
\end{equation}
 At the surface we may impose that the shape is given by a certain $\bar{\varepsilon}_s=\varepsilon_s(R)$. We will then search for the value of $\bar{\varepsilon}_s$ that causes the crust to crack (at some interior point, not only at the surface), following the Von Mises criterion.
The final thing we need, before considering the surface of the star, are the background quantities for our model, the density and the derivatives of the gravitational potential.

These calculations are all Newtonian, and thus not very accurate in the core where we expect relativistic effects to be important. As we are imposing that the core is not perturbed we can integrate the TOV equations in the core, and then match them to the Newtonian quantities we use in the crust (at the low densities of the crust the Newtonian approximation is reasonably accurate). This is important as it will allow us to use a realistic equation of state for the core, which would be pointless if we were considering a fully Newtonian star. It is in fact well known that stellar models with the same EOS and central density can give very different results in the Newtonian and relativistic cases.

 As is obvious from the above equations, both the background quantities $\rho$ and $\mu$ vanish at the surface of the star, thus causing the tractions to trivially vanish. We shall then need to impose regularity conditions.
Starting from our background model we can expand our equations around $r=R$ (the surface) and thus obtain
\begin{eqnarray}
\frac{d^2\xi_r}{dr^2}&=&\frac{1}{(r-R)}\left[\frac{\xi_r}{R}-\frac{\beta}{2}\frac{\xi_{\perp}}{R}-\frac{d\xi_r}{dr}\right]+O(1)+O(r-R)+..\\
\frac{d^2\xi_\perp}{dr^2}&=&-\frac{1}{(r-R)}\left[\frac{\xi_{\perp}}{R}+\beta\frac{\xi_r}{R}+\frac{d\xi_{\perp}}{dr}\right]+O(1)+O(r-R)+..
\end{eqnarray} 
We now have two second order equations for $\xi_r$ and $\xi_\perp$ which have decoupled from the equation for $\psi$ (which we are not interested in, as we already know what boundary condition to use for $\psi$).
The solution involves a combination of Bessel functions, this allows us to discard the ones that diverge at $r=R$ (the Bessel functions of the second kind) and thus be left with a solution described by two constants, $A_1$ and $A_2$. Expanding around $r=R$ we have:
\begin{eqnarray}
\xi_r&=&f_1(A_1,A_2)+g_1(A_1,A_2)(r-R)+O\left((r-R)^2\right)\\
\xi_\perp&=&f_2(A_1,A_2)+g_2(A_1,A_2)(r-R)+O\left((r-R)^2\right)
\end{eqnarray}
This allows us to impose relations between the functions and their derivatives close to the surface:
\begin{eqnarray}
&&\frac{d\xi_r}{dr}=-\frac{\xi_{\perp}\beta-2\xi_r}{2R}-\xi_r\frac{\beta^2+2}{2R^2}(r-R)+O\left((r-R)^2\right)\\
&&\frac{d\xi_\perp}{dr}=-\frac{\xi_\perp+\xi_r\beta}{R}-\xi_{\perp}\frac{\beta^2+2}{2R^2}(r-R)+O\left((r-R)^2\right)
\label{boundreg}\end{eqnarray}
which is the boundary condition we impose close to the surface, together with the magnitude of the perturbation, i.e. the value $\bar{\varepsilon_s}$. We thus have a total of three conditions at the surface and three at the base of the crust.

\subsection{Perturbations of the core}

So far we have derived boundary conditions for the problem of an unperturbed core with a perturbed crust. We shall now turn our attention to the more physical problem of a perturbed crust with a perturbed core. 
Let us solve the whole set of equations in the core and the crust. In the core we need to solve the same set of equations as in the crust, but now we have no shear modulus, and so they reduce to simple hydrostatic equations
\begin{equation}
\nabla p=-\rho\nabla\Phi
\end{equation}
which we can expand along our basis vectors
\begin{eqnarray}
&&\hat{r}^aY_{lm}\left(\partial_r\delta p+\rho\partial_r\delta\Phi+\delta\rho\partial_r\Phi\right)=0\\
&&\nabla^aY_{lm}\left(\delta p+\rho\delta\Phi\right)=0
\end{eqnarray}
The equations simplify to
\begin{eqnarray}
&&\delta p = -\rho\delta\Phi\\
&&\delta \rho = \frac{\partial_r\rho}{\partial_r\Phi}\delta\Phi
\label{core}\end{eqnarray}
We can substitute these into Poisson's equation
\begin{equation}
\nabla^2\delta\Phi=4\pi G\delta\rho
\end{equation}
and obtain an ODE for $\delta\Phi$
\begin{equation}
\frac{d^2\delta\Phi}{dr^2}+\frac{2}{r}\frac{d\delta\Phi}{dr}-\frac{6}{r^2}\delta\Phi=4\pi G \frac{\partial_r\rho}{\partial_r\Phi}\delta\Phi
\end{equation}
which we can solve together with equations (\ref{core}). The only boundary conditions we need to impose are those of regularity at the centre, i.e. that the solution close to $r=0$ is of the form $Ar^2$ where $A$ is a constant. This leads to
\begin{equation}
\frac{d\delta\Phi}{dr}=2\frac{\delta\Phi}{r}\;\;\;\;\;\; \mbox{close to the centre}
\end{equation}
We then impose continuity of the tractions at the base of the crust.

\section{Results for an n=1 polytrope}

\begin{figure}
\centerline{\includegraphics[height=7cm,clip]{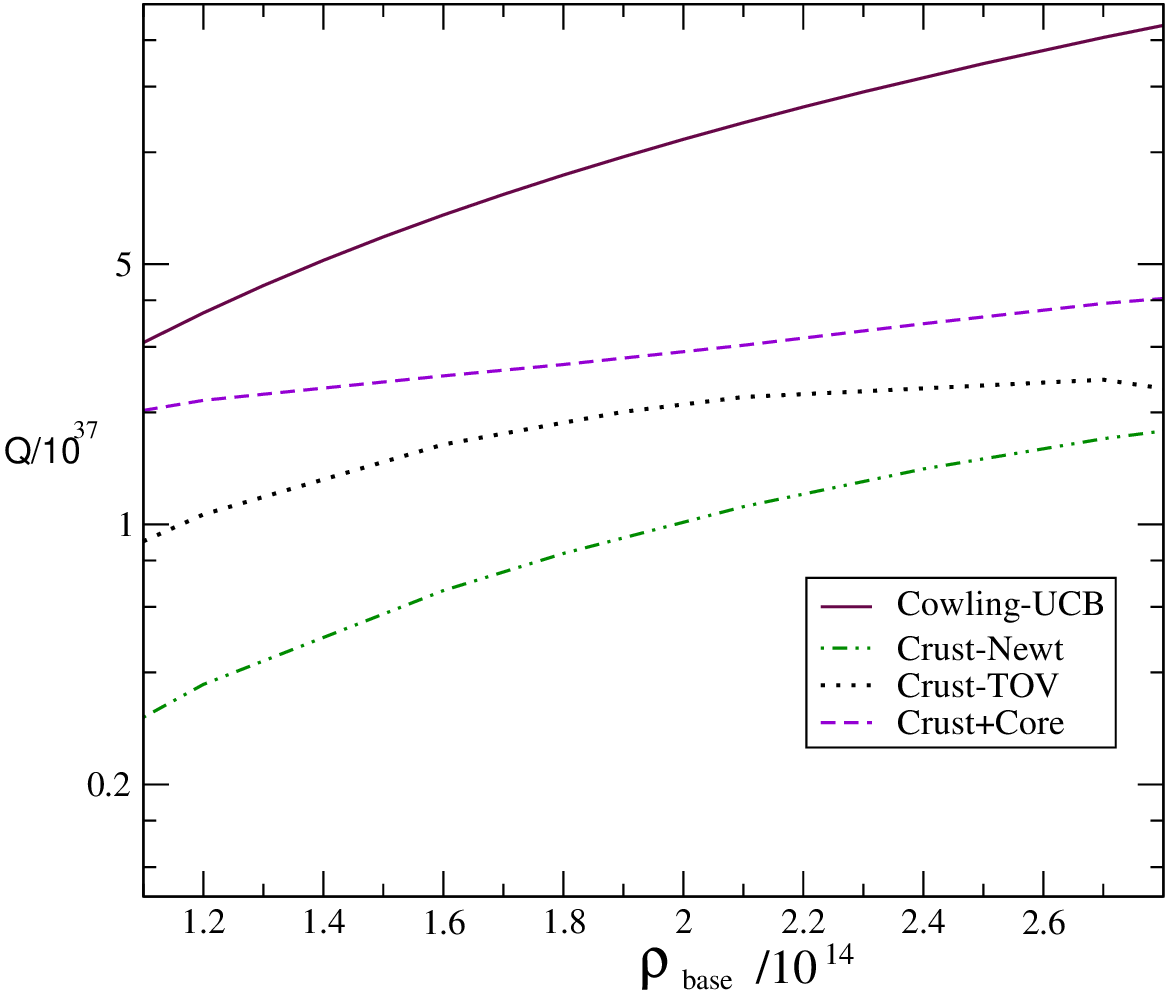}
\includegraphics[height=7cm,clip]{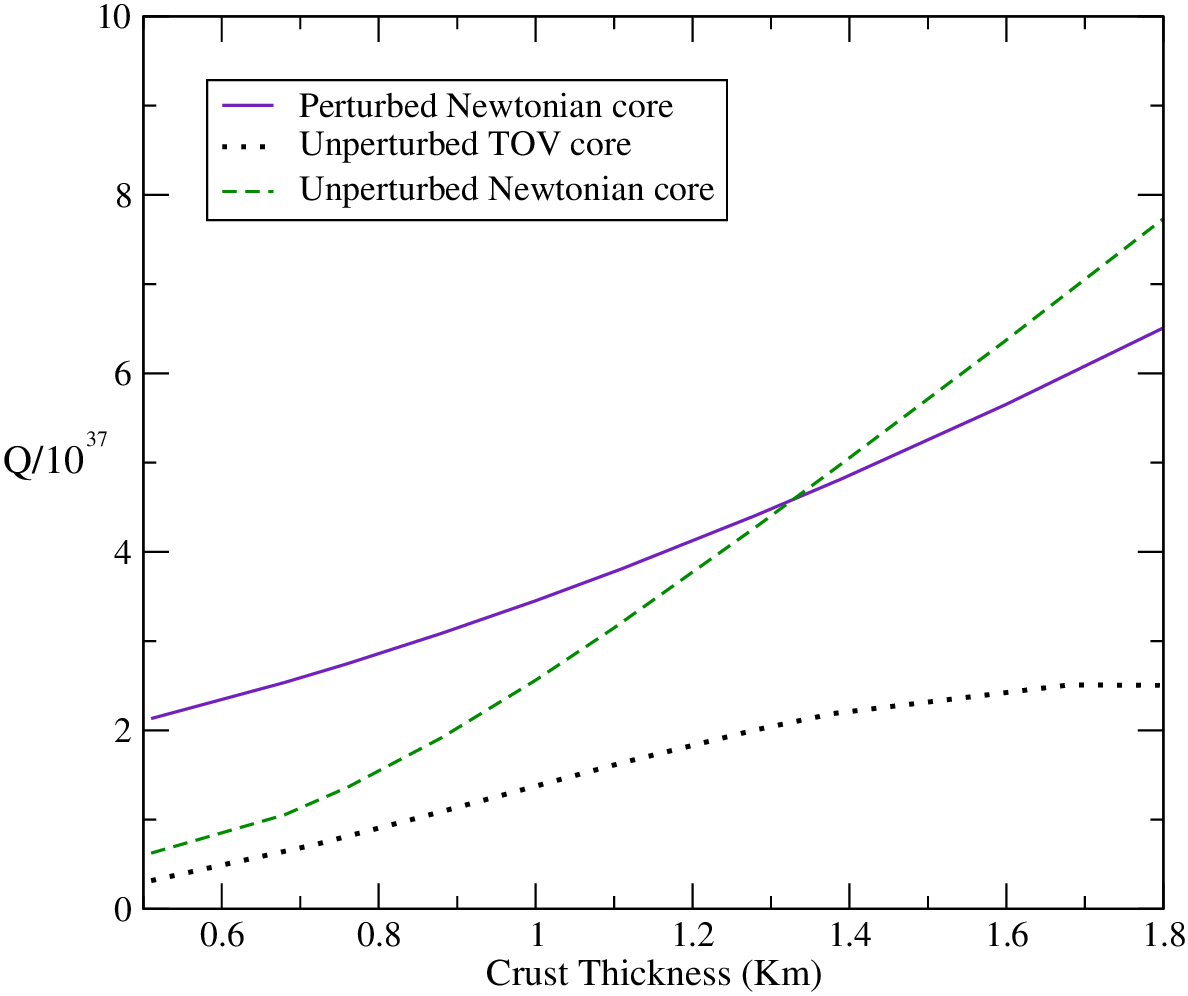} }
\caption{On the left hand side we have the maximum quadrupole plotted first with respect to the density of the base of the crust. We present the three cases described in the text and also the result of \citet{ush}, marked as UCB. On the right hand side we have the maximum quadrupole, for the three cases considered in the main text, plotted with respect to the crust thickness}
\label{thick}
\end{figure}

\begin{figure}
\centerline{\includegraphics[height=7cm,clip]{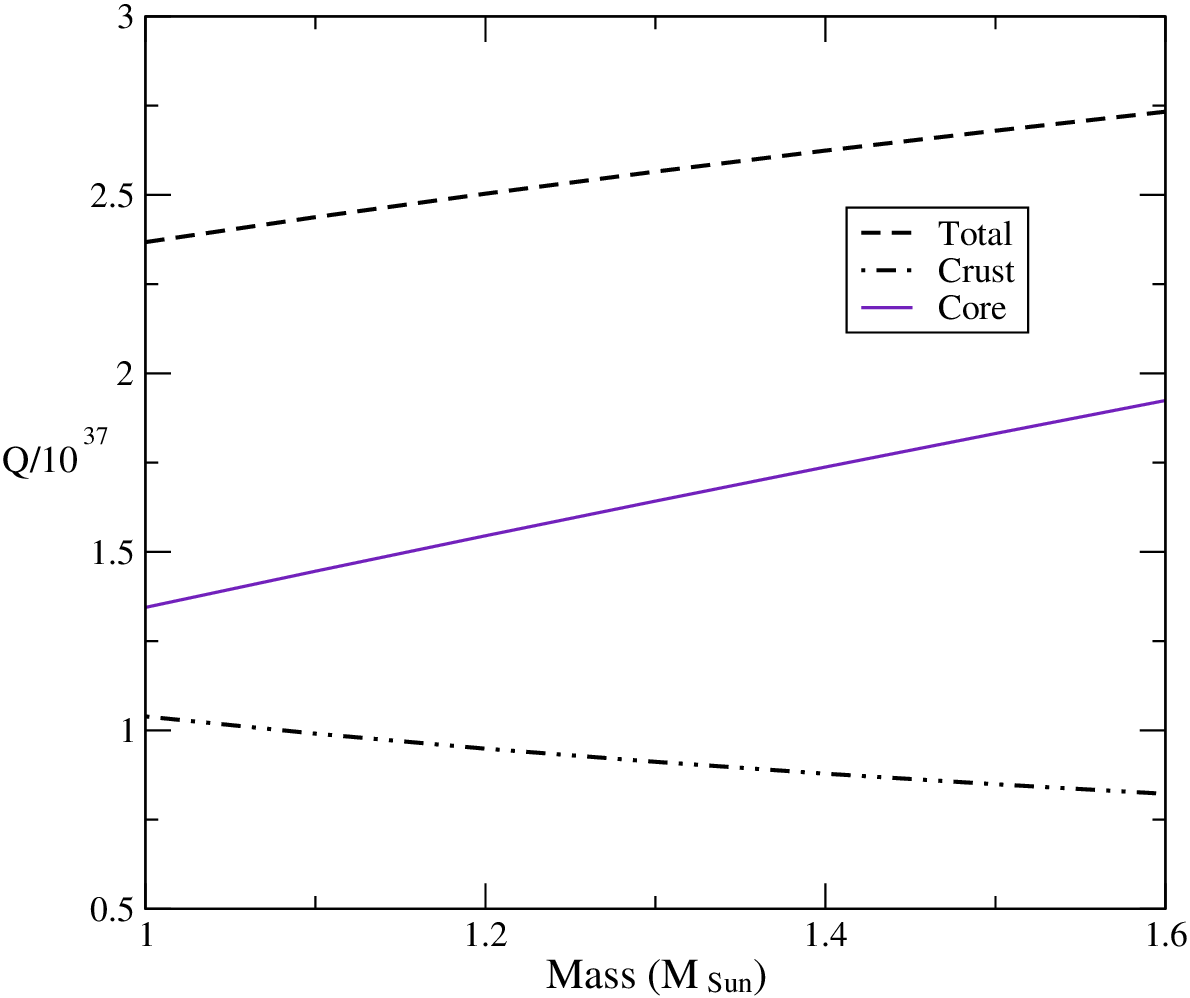}
\includegraphics[height=7cm,clip]{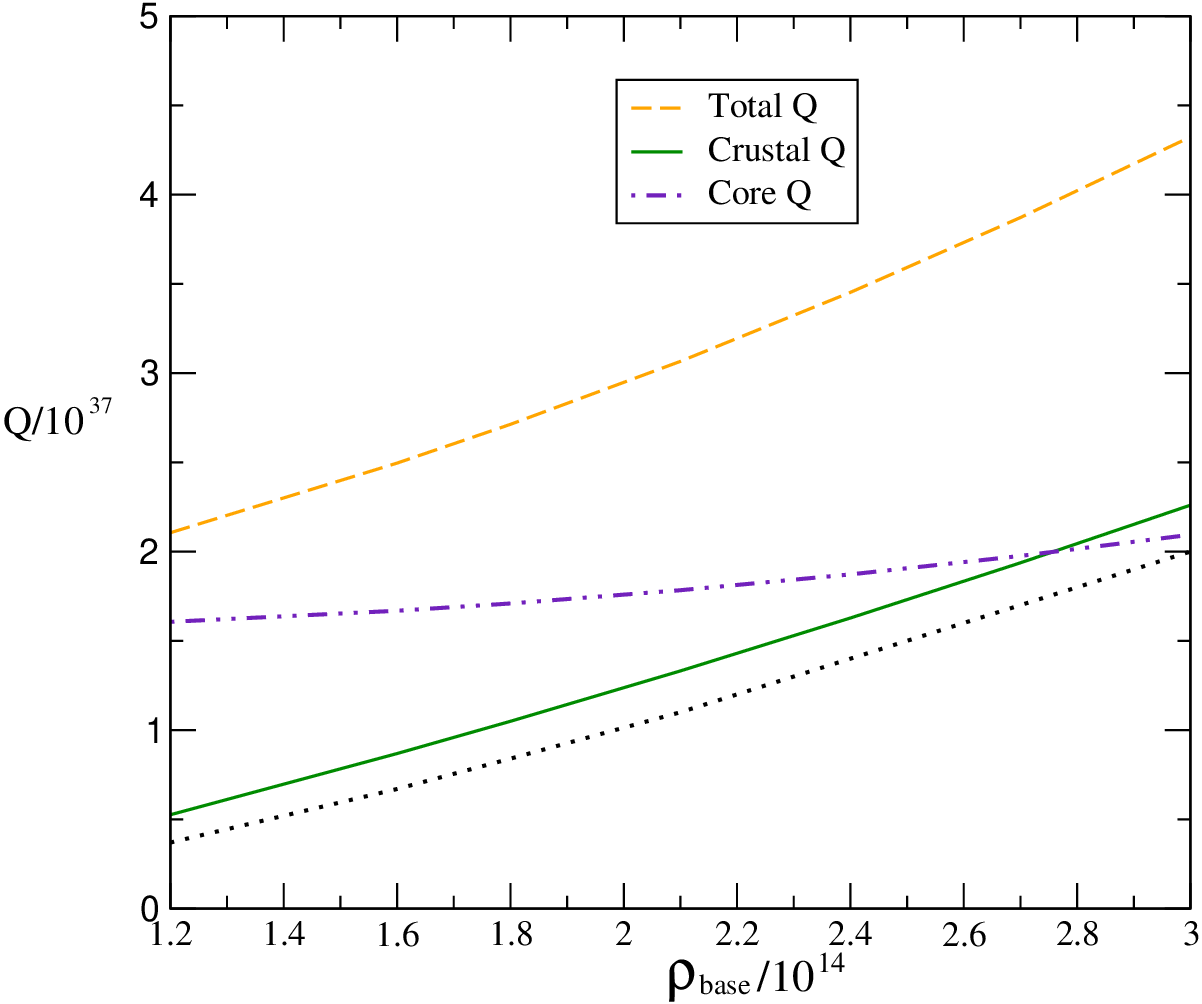} }
\caption{The dependence of the maximum quadrupole on $\rho_b$ and on the mass of the star, in the case of a perturbed core. For this case we plot also the single contributions of the crust and the core.}
\label{crustcore}
\end{figure}

\begin{figure} 
{\includegraphics[height=7cm,clip]{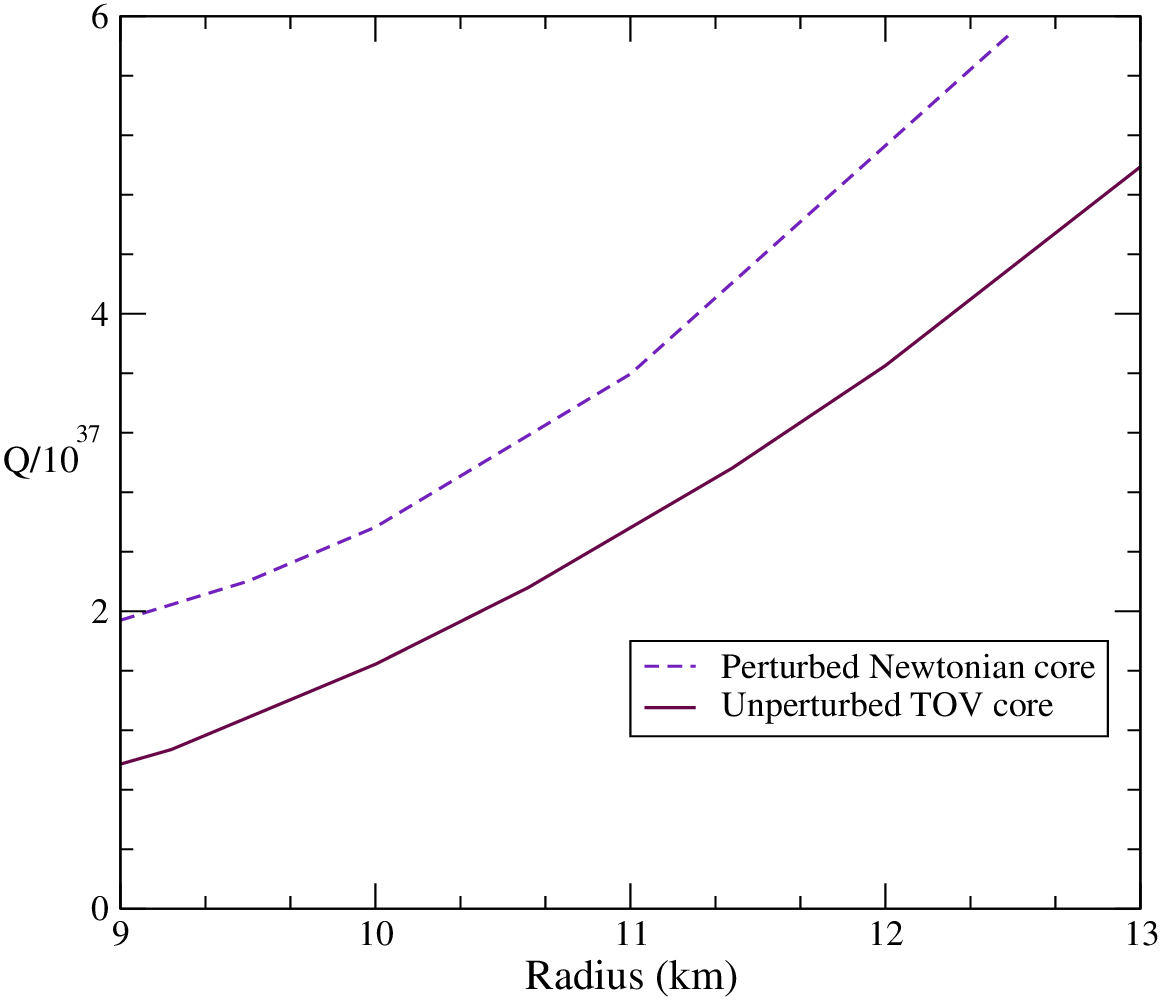}}
\caption{Dependence of the maximum quadrupole on the stellar radius, in the case of a perturbed Newtonian core and of an unperturbed TOV core.} 
\label{polyres}
\end{figure}

Let us consider a specific stellar model for which we take an $n=1$ polytropic equation of state $p=K\rho^2$, which leads to
\begin{equation}
\delta p=2K\rho\delta\rho
\end{equation}
As in previous examples the shear modulus is taken to be such that $\mu/\rho=10^{16}\;cm^2/s^2$.
We solve the equations numerically by using a shooting method with a Runge-Kutta 5th order variable step size routine (Cash-Karp). We calculate the quadrupole moment and gradually increase the value of $\psi$ until the crust breaks (following the Von Mises criterion).
 The equations are integrated in three different cases:
\begin{enumerate}
\item{Newtonian equations throughout the star, only perturbations of the crust}
\item{TOV equations in the core, Newtonian in the crust, only perturbations of the crust}
\item{Newtonian equations throughout the star, perturbations of the core and crust}
\end{enumerate}
In the first two cases the core is unperturbed, so we impose the boundary conditions derived in section 5.1, the last case includes a perturbed core, so the boundary conditions are the ones described in section 5.2.
We choose these cases as an unperturbed core has the advantage that it can be obtained by integrating either the Newtonian equations of hydrostatic equilibrium, or the TOV equations. This allows us to estimate how important overall relativistic effects may be.
The results are presented in Figure \ref{thick} taking different densities for the base of the crust. This allows us both to compare with \citet{ush}, who take the density at the base of the crust to be $2.1\times 10^{14}$g/cm$^3$, and to use the value $1.6\times 10^{14}$g/cm$^3$, suggested by \citet{Haensel}. The results for these two particular cases are shown in Table \ref{basso}.
The Newtonian calculation with the unperturbed core can be fitted to the formula\begin{equation}
Q_{\mbox{max}}=1.1\times 10^{37} \left(\frac{\rho_{b}}{1.6\times 10^{14} g/cm^3}\right)^{2}\left(\frac{R}{10 km}\right)^{6.9}\left(\frac{M}{1.4 M_{\odot}}\right)^{-1.1}\left(\frac{\sigma_{\mbox{max}}}{10^{-2}}\right) \mbox{g cm$^2$}
\end{equation}
The dependence on $\rho_{b}$ of the calculation with the relativistic core is not well fitted by a simple power law, as we can see from figure \ref{thick}, where it is evident that for thicker crusts the maximum quadrupole is growing at a slower rate. However, if we take $\rho_b=1.6\times 10^{14}$ and $R=10$ km, the dependence on the stellar mass is well fitted by 
\begin{equation}
Q_{\mbox{max}}=1.6\times 10^{37}\left(\frac{M}{1.4 M_{\odot}}\right)^{-0.3} \mbox{g cm$^2$}
\end{equation}
The Newtonian calculation including the perturbations of the core is also very interesting. As we can see from Figure \ref{thick} the quadrupole built up in the core is predominant if we take the base of the crust at low densities, which is equivalent to having a thin crust. The crust quadrupole, on the other hand, becomes predominant for thick crusts, and is well approximated by the Newtonian calculation with the unperturbed crust, as we can see in Figures \ref{thick} and \ref{crustcore}.
It is also very instructive to observe, in the case of the perturbed core, how the maximum quadrupole depends on the mass of the star (Figure \ref{crustcore}). Unlike the other two cases the quadrupole gets larger as the star gets heavier. This is because the quadrupole from the perturbations of the core is growing as the star is made more massive, while the crustal quadrupole is in fact decreasing.
The dependence on the stellar radius is shown in Figure \ref{polyres} for the case of a perturbed Newtonian core and that of an unperturbed relativistic core; the two cases are similar and not altogether surprising.
All calculations are linear in the breaking strain.

\begin{table}
\begin{tabular}{l|l|l}
&$|$Q$_{\mbox{max}}|$ (g/cm$^2$) ($\rho_b=1.6\times 10^{14}$)&$|$Q$_{\mbox{max}}|$ (g/cm$^2$) ($\rho_b=2.1\times 10^{14}$) \\
\hline
Core\&Crust-Newtonian&2.5$\times 10^{37}$($\epsilon=5.1\times 10^{-8}$)&$3.1\times 10^{37}$($\epsilon=6.3\times 10^{-8}$)\\
\hline
Crust-Newtonian &1.1$\times 10^{37}$($\epsilon=2.2\times 10^{-8}$)&1.9$\times 10^{37}$($\epsilon=3.9\times 10^{-8}$)\\
\hline
Crust-TOV+Newtonian &1.6$\times 10^{37}$($\epsilon=3.3\times 10^{-8}$)&2.2$\times 10^{37}$($\epsilon=4.6\times 10^{-8}$)\\
\hline
UCB(Newt)&6.8$\times 10^{37}$($\epsilon=1.4\times 10^{-7}$)&1.2$\times 10^{38}$($\epsilon=2.5\times 10^{-7}$)\\
\hline
\end{tabular}\caption{Maximum quadrupole, in the case of $\sigma_{\mbox{max}}=10^{-2}$ for an n=1 polytrope, and the base of the crust at $\rho_b=1.6\times 10^{14}$ g/cm$^3$ and $\rho_b=2.1\times 10^{14}$ g/cm$^3$. The parameters of the stellar models are given in table \ref{par1}.(UCB indicates the result of \citet{ush}).}\label{basso}
\begin{tabular}{l|lllll}
&M$_{core}$ ($M_{\odot}$)&crust thickness (km)&K (g$^{-1}$cm$^5$s$^{-2}$)&$\rho_c$ (g/cm$^3$)& $\rho_b$ (g/cm$^3$)\\
\hline
TOV+Newt&1.358&$1.115$&7.24$\times 10^{4}$&2.45$\times 10^{15}$&1.6$\times 10^{15}$\\
\hline
Newt&1.37&$0.68$&4.25$\times 10^{4}$&2.2$\times 10^{15}$&1.6$\times 10^{15}$\\
\hline
TOV+Newt&1.33&$1.39$&7.07$\times 10^{4}$&2.54$\times 10^{15}$&2.1$\times 10^{15}$\\
\hline
Newt&1.35&$0.88$&4.25$\times 10^{4}$&2.2$\times 10^{15}$&2.1$\times 10^{15}$\\
\hline
\end{tabular}\caption{Parameters of the two stellar models, the fully Newtonian model and the relativistic model with a Newtonian crust, for a density at the base of the crust of $\rho=1.6\times 10^{14}$ g/cm$^3$ and $\rho=2.1\times 10^{14}$ g/cm$^3$}\label{par1}
\end{table}

It would appear from the results of our calculations that considering the perturbations of the core can make a difference in the result, mainly for thin crusts. On the other hand it appears that if the crust is thick, then the crustal quadrupole plays a predominant role, and is well approximated by taking a perturbed crust on top of an unperturbed core.
In this case relativistic effects can become important, and it would then not seem appropriate to use realistic equations of state, but still describe the core with Newtonian equations of equilibrium. On the other hand it would seem to be the case that one can consider an unperturbed relativistic core (with a realistic EOS), then  use a realistic EOS and $\mu$ in the Newtonian crust and still have a reasonable approximation to the full problem. We are of course still solving the problem in Newtonian elasticity, so not solving the problem in ``full'' GR, but at the densities of the crust the Newtonian approximation is a reasonable one, and makes the problem considerably simpler than it would be in relativistic elasticity.

\subsection{Application to the Accreted vs. Non-Accreted case} 
 
We can now consider the stellar models presented in Table \ref{comparo1} for two stars, one with a non-accreted crust and one with an accreted crust. We shall first take the two stars to have the same mass,then we shall also consider two stars with the same crust thickness.
As the crust is now thick, as shown in Table \ref{comparo1}, we expect the crustal quadrupole to dominate and can thus consider the approximation of an unperturbed core to be reasonable.
This then allows us to use the equation of state given by \citet{Duc} and solve the TOV equations in the core and the Newtonian equations of hydrostatic equilibrium in the crust, thus taking relativistic effects into account.
The results are shown in Table \ref{nuova1}.
As expected the values for the quadrupole do not differ significantly. We also confirm that a neutron star with an non accreted crust could provide a slightly better source of gravitational waves, as was expected from the literature (\citet{sato}).
The values of the quadrupole (or equivalently the ellipticity) are now larger than in the previous case as we are taking a realistic model for the shear modulus (as shown in figure \ref{mu}). This leads to values of $\mu$ which are significantly larger than in the constant ratio $\mu/\rho$ approximation, especially in the higher density regions close to the base of the crust, which give rise to most of the quadrupole.
\begin{table}
\begin{tabular} {l |l |l |l}
&Accreted crust&Non Accreted crust & Accreted crust\\
\hline
Mass& $1.4 M_{\odot}$& $1.4 M_{\odot}$& $1.6 M_{\odot}$\\
Radius&$12.56$km&$12.3$km&$12.3$km\\
Crust Thickness&$1.76$km&$1.5$km&$1.5$km\\
Q$_{\mbox{max}}$&$1.8\times 10^{39}\left(\frac{\sigma}{10^{-2}}\right)$g cm$^2$&$3.1\times 10^{39}\left(\frac{\sigma}{10^{-2}}\right)$g cm$^2$&$1.6\times 10^{39}\left(\frac{\sigma}{10^{-2}}\right)$g cm$^2$\\
$\epsilon$& 1.3$\times 10^{-6}$&2.4$\times 10^{-6}$&1.1$\times 10^{-6}$
\end{tabular}\caption{Maximum quadrupole for two stars of equal mass, one with an accreted crust and one with a non-accreted crust. The last column shows a star with an accreted crust of the same thickness as that of the non accreted crust, but a different mass. The perturbative formalism of section 5 is used and the base of the crust is taken to be at $1.3\;10^{14}$g/cm$^2$ as in \citet{Duc}.}\label{nuova1} 

\end{table}

\begin{figure}
\bigskip
\begin{center}
\bigskip
\includegraphics[width = 8 cm, height=5.8 cm]{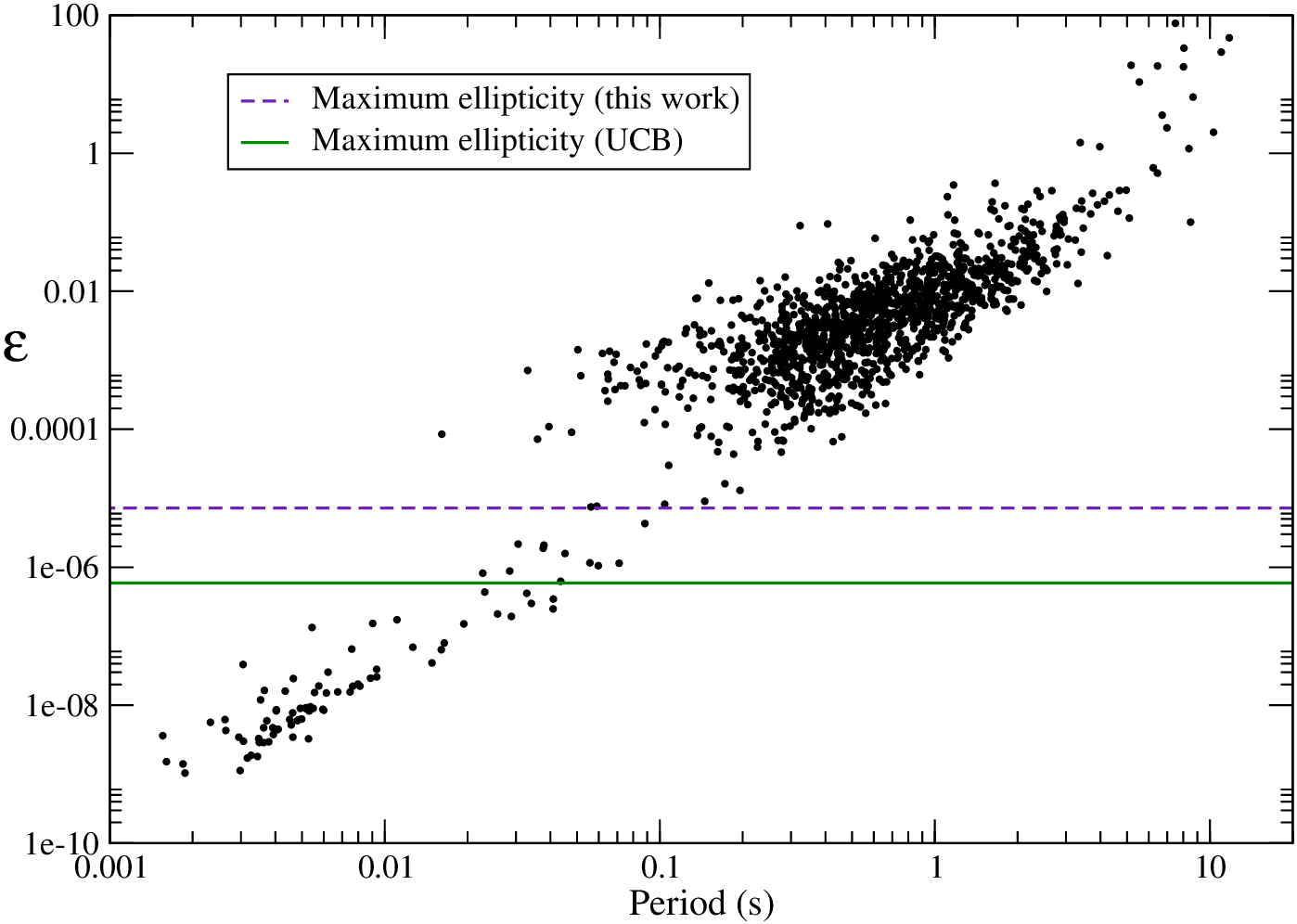}
\end{center}
\caption{The upper limits on the ellipticity $\epsilon$ set by pulsar spin down (taken from the ATNF pulsar database, www.atnf.csiro.au/research/pulsar/psrcat/) and the thoretical upper limits set by the perturbative method this work (top line) and with the method of \citet{ush} (bottom line).}
\label{spindown}
\end{figure}

The values we find for $\epsilon$ are now comparable with the upper limits of
$\epsilon\approx 10^{-5}$ set by LIGO for some of the closer pulsars (\cite{ligo}).
The best limits, however, are still set by pulsar spindown, and can be seen in Figure \ref{spindown}, compared to the limits set by this work.

\section{Conclusions}

We have reviewed the method of \citet{ush} to determine the maximum deformation that a Neutron star crust can sustain. This procedure has then been applied to the case of an accreted and a non-accreted crust, in order to estimate whether one of the two would produce a better source of gravitational waves. It appears that the two kinds of crust are essentially equivalent, with the non-accreted crust being able to support a slightly larger quadrupole, as has been suggested in the literature (e.g. \citet{sato}).
It is important to note, however, that to obtain this result is has been necessary to extrapolate the composition of the crust in the high density regions at the base of the accreted crust. In order to confirm this result it would therefore be necessary to have a precise model for the composition in this region.

We have then examined a series of toy models, to determine what effect the Cowling approximation can have on the problem and, more crucially, if it is safe to assume that the shear modulus $\mu\longrightarrow 0$ at the base of the crust.  
We conclude that this is not the case, as the shear modulus is expected to vanish over a thin transition region close to the base of the crust, which can lead to large boundary terms. The Cowling approximation can also make a difference of factors of a few in the final result.

In order to drop the Cowling approximation, and have better control over the boundary conditions, we have developed a perturbation formalism and applied it to the case of an $n=1$ polytrope with a constant ratio $\mu/\rho$. The results of these tests have confirmed that the Cowling approximation used in \citet{ush} can have a large impact. Our results, however, also suggest that it is not a bad approximation to consider an unperturbed core and perturb the crust, if the crust is sufficiently thick. This allows us to solve relativistic equations in the core, which can give corrections that are significant if we intend to use a realistic equation of state for the star.
Having developed the formalism we apply it to the original problem of the accreted and non-accreted crusts, with a realistic model for $\mu$, thus confirming that the two crusts can sustain very similar quadrupoles, the non accreted crust being able to sustain a slightly larger ``mountain''.
What is surprising is the fact that the quadrupole found with this formalism is approximately an order of magnitude greater than the results obtained with the formalism of \citet{ush}. Such a significant quadrupole, if it were confirmed, may allow us to put an upper limit on the breaking strain of the crust from observations. The problem, however, is not quite so simple. The calculations we have presented give a ``maximum'' quadrupole in as much as we have assumed all the strain to be in the $Y_{22}$ harmonic. But the stars we consider are rotating and must therefore be rotationally deformed, thus putting strain also in the $Y_{20}$ harmonic, as spin-up or spin-down moves the star away from its reference shape. This could bring the crust much closer to the yield point and significantly decrease the size of any additional ``mountain'' that it could sustain. Unfortunately, estimating exactly how much strain is built up by rotation is a complicated problem, as the unstrained configuration of the crust depends on the history of the crust, and there may have been many starquakes during the spin-down and subsequent spin-up phases.
However, if a neutron star can sustain a quadrupole this large for a breaking strain of $\sigma_{\mbox{max}}=10^{-2}$, it would appear that even if the breaking strain is smaller there is still a possibility that the quadrupole may be large enough for gravitational waves to have a role in the spin evolution of the star.

Finally we should note that our approach was chosen in order to obtain a ``maximization'' argument, independent of the mechanism that produces the deformation, and to show that overall relativistic effects can be important.
We have, therefore, integrated a set of source free equations, analogous to those of \citet{mac}, with the deformation as a boundary condition. It would, however, be important to understand the physical processes that cause the deformation and  integrate the same equations with a source, such as the temperature or composition variations used in \citet{ush}.
Another important effect we are neglecting is that of the magnetic field. It is well known that the equilibrium shape of a star will not be spherical in the presence of a magnetic field and, even though crustal effects may be dominant early on during accretion, the process of magnetic burial could make the field locally strong enough for the magnetic deformations to become important. This effect was recently studied by \citet{melatos} and \citet{payne}, who were the first to perform globally self-consistent MHD calculations of magnetic mountains on accreting NS as gravitational wave sources. These calculations did not include crustal effects (as this would make the problem much less tractable), nevertheless it is important to note here that if one were to solve the full inhomogeneous problem, including the magnetic source terms (or terms due to other physical process we have neglected), one could obtain deformations larger than the crustal maximum we have calculated.
 
It would also be of some interest to consider the full problem in a relativistic formulation of elasticity, such as that of \citet{lars}, as we have shown that the perturbations of the core can be important, especially in determining the dependence of the maximum quadrupole on the stellar parameters.
We intend to turn our attention to these problems in the near future.


\end{document}